\documentclass[aps,prd,onecolumn,nofootinbib,superscriptaddress]{revtex4}
\pdfoutput=1
\usepackage{float}
\usepackage{graphicx}%Include figure files
\usepackage{amsmath}
\usepackage{amsfonts}
\usepackage{amssymb,ulem}
\usepackage{color}%
\usepackage{dcolumn}
\usepackage{subfigure}
\usepackage{multirow}
\usepackage{MnSymbol,wasysym}
\usepackage{braket,diagbox}
\usepackage{eurosym}
\usepackage{calrsfs}
\usepackage[usenames,dvipsnames,svgnames]{xcolor}

\newcommand{\blue}{\textcolor[rgb]{0.33,0.33,1.00}}

\newcommand{\RNum}[1]{\uppercase\expandafter{\romannumeral #1\relax}}
\usepackage[colorlinks=true,linkcolor=blue,urlcolor=blue,filecolor=black,citecolor=red,
pdfstartview=FitV,pdftitle={},pdfsubject={},pdfkeywords={},pdfpagemode=None,bookmarksopen=true]{hyperref}

\usepackage{pifont}

\begin{document}
\baselineskip=0.4 cm
%%%%%%%

\title{Images from disk and spherical accretions of hairy Schwarzschild black holes}
%%%%%%%
\author{Yuan Meng}
\email{mengyuanphy@163.com}
\affiliation{Center for Gravitation and Cosmology, College of Physical Science and Technology, Yangzhou University, Yangzhou, 225009, China}

\author{Xiao-Mei Kuang}
\email{xmeikuang@yzu.edu.cn (corresponding author)}
\affiliation{Center for Gravitation and Cosmology, College of Physical Science and Technology, Yangzhou University, Yangzhou, 225009, China}

\author{Xi-Jing Wang}
\email{xijingwang@yzu.edu.cn}
\affiliation{Center for Gravitation and Cosmology, College of Physical Science and Technology, Yangzhou University, Yangzhou, 225009, China}

\author{Bin Wang}
\email{wang$_$b@sjtu.edu.cn}
\affiliation{Center for Gravitation and Cosmology, College of Physical Science and Technology, Yangzhou University, Yangzhou, 225009, China}
\affiliation{Shanghai Frontier Science Center for Gravitational Wave Detection, Shanghai Jiao Tong University, Shanghai 200240, China}
\author{Jian-Pin Wu}
\email{jianpinwu@yzu.edu.cn}
\affiliation{Center for Gravitation and Cosmology, College of Physical Science and Technology, Yangzhou University, Yangzhou, 225009, China}

\date{\today}

\begin{abstract}
\baselineskip=0.5 cm
A hairy Schwarzschild black hole describes the deformation of Schwarzschild black hole due to including additional sources. It is found that depending on the hairy parameters, the photons' configurations around this black hole can be classified  into two cases, corresponding to the hairy Schwarzschild black hole with single photon sphere and double photon spheres, respectively. We focus on the shadows and images of the hairy Schwarzschild black hole under two types of static thin illuminations conditions: disk accretion and spherical accretion, respectively. Under both illuminations, the two hairy parameters ($\alpha$ and $l_o$) have competitive effects on the shadow and optical appearance image of the hairy Schwarzschild black hole with single photon sphere. This means that even though the parameters have significant influences on the rings and shadows, its images with certain groups of $\alpha$ and $l_o$ could be indistinguishable to that of Schwarzschild black hole, namely, the degeneracies of images raise between the hairy Schwarzschild black hole and Schwarzschild black hole. Moreover, the optical appearance image of the hairy Schwarzschild black hole with double photon spheres will exhibit new additional rings and accretion features, which {are not} present in the images of (hairy) Schwarzschild black hole with single photon sphere. Our theoretical studies on the rings and shadows provide a potential tool to differentiate the hairy Schwarzschild black hole with double photon spheres from Schwarzschild black hole, but they are not always helpful for the cases with single photon sphere due to the degeneracies.
\end{abstract}

%%%%%%%

\maketitle
\newpage
\tableofcontents

\section{Introduction}
 Recent breakthroughs in the observation of black holes have triggered a new era to approach strong gravity field regime to further testify the essence of gravity. One of the most important
achievements is that the Event Horizon Telescope (EHT) collaboration has released images of the supermassive black holes in M87* \cite{EventHorizonTelescope:2019dse,EventHorizonTelescope:2019uob,EventHorizonTelescope:2019jan,EventHorizonTelescope:2019ths,EventHorizonTelescope:2019pgp,EventHorizonTelescope:2019ggy}, and further in Sgr A* at the center of the Milky Way system \cite{EventHorizonTelescope:2022wkp,EventHorizonTelescope:2022apq,EventHorizonTelescope:2022wok,EventHorizonTelescope:2022exc,EventHorizonTelescope:2022urf,EventHorizonTelescope:2022xqj}.
Those images  show a black central region surrounded by a bright ring-shaped construction, which is the resultant product of light
rays  in the gravitational field of an object having a photon sphere (or a critical unstable curve) when illuminated
by an accretion flow.  The central silhouette which is bounded by the critical curve is usually known as the black hole shadow or photon capture region \cite{Perlick:2021aok}.

Technically, the critical curve is defined as the light ray received by the observer, that
would have approached asymptotically a bound photon orbit in ray traced backwards methods.
In fact, early on, Synge and Luminet proposed the expression of the angular radius of the photon capture region for Schwarzschild black holes, which is determined by a critical impact parameter \cite{Synge:1966okc,Luminet:1979nyg}. Subsequently, Bardeen
presented the shadow of the rotating Kerr black hole for the first time, indicating that the spin can deform the black hole shadow \cite{Bardeen:1973}. Due to the fact that the black hole shadow only depends on the background geometry,  various numerical simulations of shadows in general relativity (GR) have been extensively discussed \cite{Falcke:1999pj,Shen:2005cw,Yumoto:2012kz,Atamurotov:2013sca,Papnoi:2014aaa,Abdujabbarov:2015xqa,Kumar:2018ple}. In addition,  the black hole shadows in various modified theories of gravity (MoG) and in high-dimensional space-time have been extensively studied \cite{Amarilla:2010zq,Amarilla:2011fx,Amarilla:2013sj,Amir:2017slq,Singh:2017vfr,Mizuno:2018lxz,Eiroa:2017uuq,Vagnozzi:2019apd,Banerjee:2019nnj}. Moreover, though the discovery from EHT are mainly based on black holes in GR, it allows plenty of room for alternative compact objects or black holes in theories beyond GR.
So, using the EHT observations of the shadow to testify fundamental physics and constrain parameters in MoG is far reaching, see for examples \cite{Afrin:2021imp,Meng:2022kjs,Kuang:2022ojj,Kuang:2022xjp,Tang:2022hsu,Tang:2022bcm,Kumar:2019pjp,Kumar:2020hgm,Shaikh:2021yux,Meng:2023wgi,Wu:2023yhp,Ghosh:2023kge} and references therein.

The bright ring-shaped construction in the EHT images is radiated from the  accretion matters surrounding the real astrophysical black holes,  of which the geometry and physical properties could significantly determine the optical appearance of the black hole. In theoretical aspect, it is difficult to mimic the realistic accretion flow in astrophysical environment, but it is useful to consider some simplified accretion conditions to investigate the major features of black hole image and capture prospective signal of new physics.  The first image of black hole with a thin accretion disk was calculated analytically in \cite{Luminet:1979nyg}, which shows that there are primary and secondary images appeared outside black hole shadow.  Then  the author of  \cite{Bambi:2013nla}  pointed out that it is relatively easy to distinguish a Schwarzschild black hole from the static wormhole according to shadow images.  As another kind of accretion, the spherical accretion has been applied to analyze the image of a Schwarzschild black hole \cite{Falcke:1999pj,Narayan:2019imo} in which the shadow is found to be a robust feature, and  its size and shape are primarily influenced by rather the spacetime geometry than the  details of the accretion. More recent investigations of a Schwarzschild black hole with thin and thick accretion disks \cite{Gralla:2019xty} show that the lensed ring together with photon ring contribute additional observed flux to the image, but the main contribution of the total observed specific intensities comes from direct emissions, while the contribution of the lensed ring emissions is small, and the contribution of the photon ring emissions is negligible.
{Nowadays, the photon ring and observational appearances of black holes is an exciting area of research. They allow one to differentiate GR black holes from alternative compact objects \cite{Boshkayev:2022vlv,Xavier:2023exm,Sakai:2014pga,Bacchini:2021fig,Destounis:2023khj} or black holes in MoG \cite{Zeng:2020dco,Saurabh:2020zqg,Zeng:2020vsj,Qin:2020xzu,Guo:2022iiy,Archer-Smith:2020hqq,Dokuchaev:2019pcx,Peng:2020wun,Okyay:2021nnh,Eichhorn:2021iwq,Li:2021riw,Guo:2021bhr,Zeng:2022pvb,He:2021htq,He:2022aox,Gan:2021pwu,
Gussmann:2021mjj,Hou:2022eev,Uniyal:2022vdu,Uniyal:2023inx,Akbarieh:2023kjv,Wang:2023vcv} via illumination,  because their respective images can be used to distinguish compact objects in GR and beyond.
However,  the optical features do not always work in the distinguishing scenario because   the degeneracies may arise, even in GR.  For instance, the Boson star \cite{Liebling:2012fv}, as  an alternative to black hole within GR, could  produce conventional black hole shadow-like images if  their compactness is large enough \cite{Rosa:2022tfv}.  It was also found in \cite{Vincent:2015xta} that very relativistic rotating boson stars at the Galactic center could have images extremely similar to Kerr black holes. Additionally, it was addressed in \cite{Vincent:2015xta,Cardoso:2022whc,Destounis:2022obl,Figueiredo:2023gas} that in Einstein's GR,  the addition of an astrophysical environment around a black hole can make it dress external `hair', which could bring in  similar optical features to the  Schwarzschild one even though the degeneracies may break in certain high compactness regime \cite{Boshkayev:2022vlv,Xavier:2023exm}.   }

%\Meng{It is worth noting that the addition of an astrophysical environment around a black hole \cite{Barausse:2014tra,Cardoso:2021wlq,Boshkayev:2022vlv,Cardoso:2022whc,Destounis:2022obl,Figueiredo:2023gas,Xavier:2023exm}, or a solution of general relativity that is not a black hole but rather an exotic compact object\cite{Liebling:2012fv,Sakai:2014pga,Vincent:2015xta,Bacchini:2021fig,Rosa:2022tfv,Destounis:2023khj}, can lead to external "hair" without necessarily deviating from Einstein's theory at all.}

On the other hand, due to the additional surrounding sources, the black holes in our Universe could obtain an extra global charge dubbed `hair’ and the spacetime may deviate from the black hole metric in GR.
Recently, a hairy Schwarzschild  black hole was constructed with the use of the gravitational decoupling (GD) approach  \cite{Ovalle:2020kpd,Contreras:2021yxe}, which is designed for describing deformations of known solutions of GR due to the inclusion of additional sources. The GD approach and the metric will be reviewed soon in next section. The  hairy Schwarzschild black hole and its rotating counterpart have attracted quite a lot {of} attentions. Plenty of  theoretical and observational investigations have been studied, for examples,  thermodynamics \cite{Mahapatra:2022xea}, quasinormal modes and (in)stability \cite{Cavalcanti:2022cga,Yang:2022ifo,Li:2022hkq},  strong gravitational lensing, parameter constraint from EHT observations on black hole shadow \cite{Islam:2021dyk,Afrin:2021imp}, precession and Lense-Thirring effect \cite{Wu:2023wld} and gravitational waves from extreme mass ratio inspirals \cite{Zi:2023omh}. Those investigations  promote the possible test of the no-hair theorem and could provide powerful probes of alternative theories of gravity with additional fields.

The main aim of this work is to study the rings and  optical appearances of the hairy Schwarzschild black hole proposed in GD approach. The interest mainly stems from two aspects: (i) the hairy black hole in this scenario has great generality because  there is no certain matter fields in the GD approach,  so this hairy metric allows us to study
the light rays and shadow effected by arbitrary type of hair (e.g. scalar hair, tensor hair, fluid-like dark matter, and so on)  and compare them to that of Schwarzschild black hole. (ii) The hairy Schwarzschild black hole was addressed in \cite{Guo:2022ghl} to possess two unstable photon spheres outside the event horizon in certain parameters region, therefore, it is natural to expect that the second photon sphere will bring in rich structures in the observed appearance.
Thus, we shall firstly analyze the effective potential of
the photons in the hairy parameters space, and then study the light rays around the hairy Schwarzschild black hole with the use of ray tracing method, which is significantly affected by the hairy parameters. We find that the photons' configurations in essence can be classified into two cases, corresponding to hairy Schwarzschild black hole with single photon sphere and double photon spheres, respectively. Then by illuminating the  hairy  black hole with various static thin accretions, we analyze the effects of the hairy parameters on the rings and shadows in the optical appearance images, and also differentiate the image of hairy Schwarzschild black hole with double photon spheres from that with single photon sphere.

The paper is organized as follows. In Section \ref{sec:photon and potential}, we briefly review the hairy Schwarzschild black hole constructed by GD approach and then analyze its photon spheres. In Section \ref{sec:photon trajectory}, with the use of ray tracing method, we study the light rays distributions around the hairy black hole with various parameters. We then explore the optical appearances images of the hairy Schwarzschild black hole with both single photon sphere and double photon spheres, when it is under the illumination of  static thin accretion disk (section \ref{sec:thin disk accretions}) and spherical accretion (section \ref{sec:Static spherical accretions}), respectively.  Finally, section \ref{sec:conclusion} is our closing remarks.

\section{Photon sphere of the hairy Schwarzschild black holes} \label{sec:photon and potential}
In this section, we will show a brief review on the idea of GD approach and  the hairy Schwarzschild black hole constructed from GD approach by  Ovalle \cite{Ovalle:2020kpd}. Then we investigate the nature of motions of photon in the vicinity of the hairy black hole.

The no-hair theorem in classical GR states that black holes are only described by mass, electric charge and spin \cite{Ruffini:1971bza}. But it is possible that the interaction between black hole spacetime and matters brings in other charge, such that the black hole could carry hairs. The physical effect of  these hairs can modify the spacetime of the background of black hole, namely hairy black holes may form. Recently, Ovalle et.al used the GD approach to obtain a spherically symmetric metric with hair \cite{Ovalle:2020kpd},
in which the corresponding Einstein equation is expressed by
\begin{equation}\label{eq-EE}
G_{\mu\nu}\equiv R_{\mu\nu}-\frac{1}{2}Rg_{\mu\nu}=8\pi\Tilde{T}_{\mu\nu}.
\end{equation}
 with the total energy momentum tensor $\Tilde{T}_{\mu\nu}$ containing two parts, $\Tilde{T}_{\mu\nu}=T_{\mu\nu}+\vartheta_{\mu\nu}$. Here  $T_{\mu\nu}$ is the energy momentum tensor associated with a known solution of  GR and $\vartheta_{\mu\nu}$ is  introduced by new matter fields or a new gravitational sector.  $\nabla^\mu \Tilde{T}_{\mu\nu}=0$ is satisfied because of the Bianchi identity. The main aspect of GD approach, which was originally proposed in  \cite{Ovalle:2017fgl} and used extensively by Ovalle et.al, exists in the fact that $\vartheta_{\mu\nu}$ is decoupled from  $T_{\mu\nu}$ \cite{Contreras:2021yxe,Ovalle:2020kpd}. To better understand how the GD approach works in the construction of a deformed solution, we will review the main technical points, and then demonstrate  that under the decoupling assumption, one could indeed decouple the equations of motion for the two sectors. Firstly, one can write the spherically symmetric and static solution ${g}_{\mu\nu}$ to \eqref{eq-EE} as
\begin{equation}\label{eq-swx0}
ds^2=-e^{\nu(r)}dt^2+e^{\lambda(r)}dr^2+r^2(d\theta^2+\sin^2\theta d\phi^2),
\end{equation}
 such that we have the Einstein tensor $G_{\mu}^{~\nu}(\nu(r),\lambda(r))$. Secondly, one considers that the above solution \eqref{eq-swx0} is generated by the seed metric with only the source $T_{\mu\nu}$ (i.e. $\vartheta_{\mu\nu}=0$),
 \begin{equation}\label{eq-swx1}
ds^2=-e^{\xi(r)}dt^2+e^{\mu(r)}dr^2+r^2(d\theta^2+\sin^2\theta d\phi^2),
\end{equation}
and the introduction of the source $\vartheta_{\mu\nu}$ is equivalent to the deformation of the seed metric via
\begin{gather}\label{eq-deswx}
\xi(r)\rightarrow\nu(r)= \xi(r)+\alpha~k(r), ~~~~~~~e^{-\mu(r)}\rightarrow e^{-\lambda(r)}= e^{-\mu(r)}+\alpha~h(r),
\end{gather}
where the parameter $\alpha$ keeps track of the deformations. By using the above deformations, it is easy to check that the Einstein equation \eqref{eq-EE} can be splitted into {${G}_{\mu}^{~\nu}(\xi(r),\mu(r))=8\pi T_{\mu}^{~\nu}$ and $\alpha~\mathcal{G}_{\mu}^{~\nu}(\xi(r),\mu(r);k(r),h(r))$=$8\pi\vartheta_{\mu}^{~\nu}$}, where the former is the standard Einstein equation.
Obviously, the tensor $\vartheta_{\mu\nu}$ will vanish once the metric deformations vanish ($\alpha=0$). Note that under the transformation \eqref{eq-deswx}, the Einstein tensor changes as
\begin{gather}
{G}_{\mu}^{~\nu}(\xi(r),\mu(r))\to  G_{\mu}^{~\nu}(\nu(r),\lambda(r)) ={G}_{\mu}^{~\nu}(\xi(r),\mu(r))+\alpha~\mathcal{G}_{\mu}^{~\nu}(\nu(r),\lambda(r)),
\end{gather}
which is a linear decomposition of the Einstein tensor, similar as the two sources added linearly in the r.h.s of \eqref{eq-EE}. This is the fact that makes the GD approach work.

Further, one can consider the seed metric \eqref{eq-swx1} as the Schwarzschild one, i.e, in the vacuum case with  $T_{\mu\nu}=0$. Then, by treating the additional source as the anisotropic fluid satisfying strong energy condition, one can solve out the Einstein equation and obtain the hairy solution deformed from the Schwarzschild metric. The calculations are straightforward and the readers can refer to  \cite{Ovalle:2020kpd,Contreras:2021yxe} for more details.
Here we will omit their steps, and directly write down the formula of the metric for the hairy Schwarszchild black hole
\begin{equation}\label{eq-static}
ds^2=-f(r)dt^2+\frac{dr^2}{f(r)}+r^2(d\theta^2+\sin^2\theta d\phi^2)
~~\mathrm{with}~~ f(r)=1-\frac{2M}{r}+\alpha e^{-r/(M-l_o/2)}.
\end{equation}
This metric describes certain deformation of the Schwarszchild solution due to the introduction of additional material sources, which can be scalar hair, tensor hair, fluid-like dark matter, and so on. When $\alpha=0$,  the metric reduces to the Schwarzchild solution in GR, namely with the absence of the matters.
In this solution, $M$ is the black hole mass, and $\alpha$ is the deformation parameter due to the introduction of surrounding matters, which describes the physics related with the strength of hairs. $l_o=\alpha l$, where $l$ is a parameter with length dimension, corresponds to the charge of primary hair which should satisfy $l_o\leq 2M$ to guarantee the asymptotic flatness.
The event horizon is determined by the root to the equation
\begin{eqnarray}
f(r)\mid_{r=r_h}=0\Rightarrow r_h+\alpha r_h e^{-r_h/(M-l_o/2)}=2M,
\end{eqnarray}
which can only be solved with numeric.  It is noted that the above equation has unique root, which is always positive, indicating the event horizon; while further recalling the prescription of strong energy condition requires the parameters to fulfill $r_h\geq 2M-l_o$, but its thermodynamic properties are the same as that of Schwarzschild black hole \cite{Ovalle:2020kpd}.

Next, we will build the equations for the null geodesic motion of the hairy Schwarzschild black holes. The motions of photons are described by Euler-Lagrange equation,
\begin{eqnarray}\label{eq:EL}
\frac{d}{d\lambda}\Big(\frac{\partial \mathcal{L}}{\partial \dot{x}^\mu}\Big)-\frac{\partial \mathcal{L}}{\partial x^\mu}=0,
\end{eqnarray}
where $\dot{x}^\mu=\frac{dx^\mu}{d\lambda}$ represents the four-velocity of photon with $\lambda$ the affine parameter, and for the metric \eqref{eq-static} the Lagrangian of the photon is
\begin{eqnarray}
\mathcal{L}=\frac{1}{2}g_{\mu\nu}\dot{x}^\mu\dot{x}^\nu=\frac{1}{2}\big[-f(r)\dot{t}^2+\frac{1}{f(r)}\dot{r}^2+r^2(\dot{\theta}^2+\sin^2\theta\dot{\phi}^2)\big].
\end{eqnarray}
Since $\partial_t$ and $\partial_{\phi}$ are Killing vector fields in the hairy Schwarzschild spacetime,  we can obtain two conservation constants $E$ and $L_{z}$,
\begin{eqnarray}
E\equiv-\frac{\partial \mathcal{L}}{\partial \dot{t}}=f(r)\dot{t},~~~~~~~~~~L_{z}=\frac{\partial \mathcal{L}}{\partial \mathcal{\dot{\phi}}}=r^2 \sin^2 \theta\dot{\phi},
\end{eqnarray}
which indicate the energy  and z-component of the angular momentum of the photon.
Moreover, due to the spherical symmetry of the spacetime, we can consider the motions of photon on the equatorial plane ($\theta=\pi/2$) for convenience. Then, considering  $\mathcal{L}=0$ for photon and defining the impact parameter  $b\equiv L_{z}/E$, we can extract  three equations of motion for the photons around the hairy Schwarzschild black hole from \eqref{eq:EL} as
\begin{eqnarray}
\dot{t}=\frac{1}{bf(r)},~~~~
\dot{\phi}=\pm\frac{1}{r^2},
\label{eq:phi-motion}~~~~~
\end{eqnarray}
\begin{eqnarray}
\dot{r}^2=\frac{1}{b^2}-V_{\text{eff}}(r),
\label{eq:r-motion}
\end{eqnarray}
where the effective potential takes the form
\begin{equation}
V_{\text{eff}}(r)=\frac{1}{r^2}f(r).
\end{equation}
The radial geodesic equation \eqref{eq:r-motion} determines the fate of a given photon depending on its impact parameter. In particular,
when $b$ satisfies $\frac{1}{b^2}=V_{\text{eff}}(r_0)$ at certain $r=r_0$, we have $\dot{r}=0$ which means that the photon is deflected at the minimum distance $r_0$ from the central black hole. The critical impact parameter, $b_{ph}$, is determined by the vanishing of $\dot{r}$ at the maximum of the potential,
\begin{eqnarray}
b_{ph}=\frac{1}{\sqrt{V_{\text{eff}}(r_{ph})}}, \qquad V_{\text{eff}}'(r_{ph})=0, \qquad V_{\text{eff}}''(r_{ph})<0. \label{eq:formulabrph}
\end{eqnarray}
Thus, the geodesic trajectories with $b_{ph}$ are usually known as photon spheres with radius $r_{ph}$, and the orbits of these light rays are unstable, meaning that a small radial perturbation will make it either run ($b>b_{ph}$) to infinity or to be captured ($b<b_{ph}$) into the event horizon of the black hole.

For static black hole, the photon sphere generates the boundary of black hole shadows, and plays a key role in the image of black hole. The existence of single photon sphere outside the event horizon of black hole is commonly considered because it  physically connects with the respect of dominant energy condition (DEC) and strong energy condition (SEC) \cite{Cvetic:2016bxi}. More recently, physicists have great interests in the black holes with two photon spheres, inspired by the work \cite{Liu:2019rib} in which the authors found the existence of two photon spheres outside the event horizon of dyonic black holes when the SEC is violated. It was found that double photon spheres could
lead to echo signal in dyonic black holes \cite{Huang:2021qwe} and profound observed appearances with various illuminations around scalarized RN black hole \cite{Gan:2021xdl,Guo:2022muy,Chen:2022qrw}. In particular, very recently,  many black holes were found to have two photon spheres even though DEC and SEC are satisfied \cite{Guo:2022ghl}. The hairy Schwarzschild black hole \eqref{eq-static} is the one they  considered, and double photon spheres can exist when both $\alpha$ and $l_o$  are large enough of which the parameter region were present in the figure 6 of \cite{Guo:2022ghl}.

\begin{figure} [htbp]
\centering
\subfigure[\, the radius of photon sphere(s)]
{\includegraphics[width=8cm]{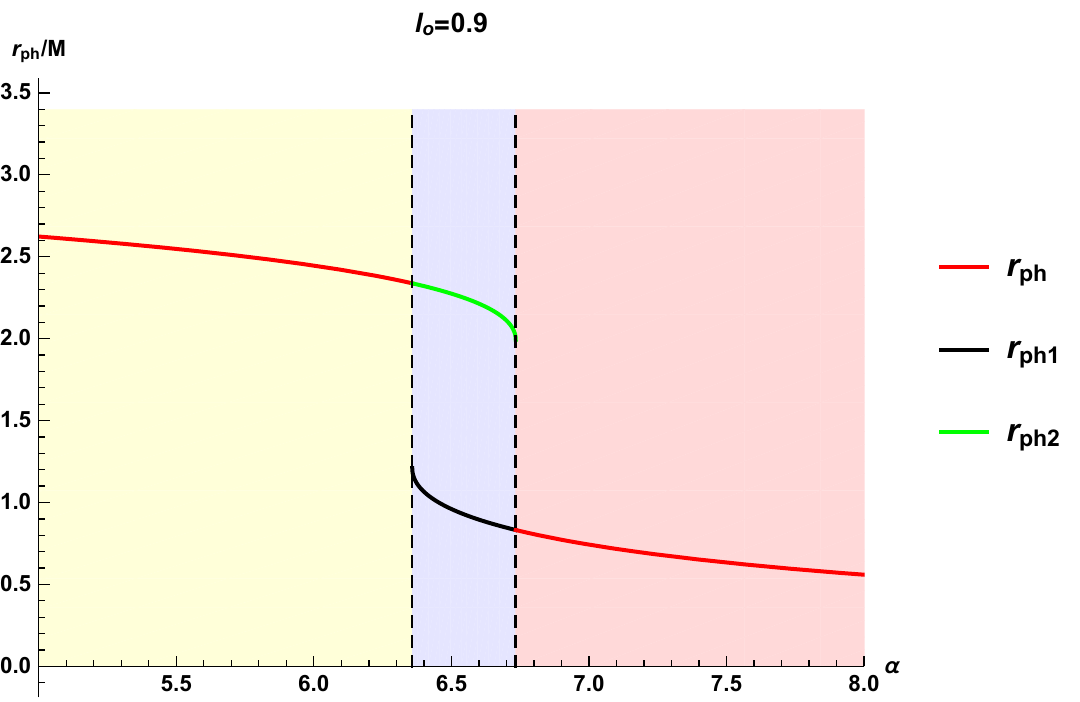}\label{fig:rp}}\hspace{1cm}
\subfigure[\, the critical impact parameter(s)]
{\includegraphics[width=8cm]{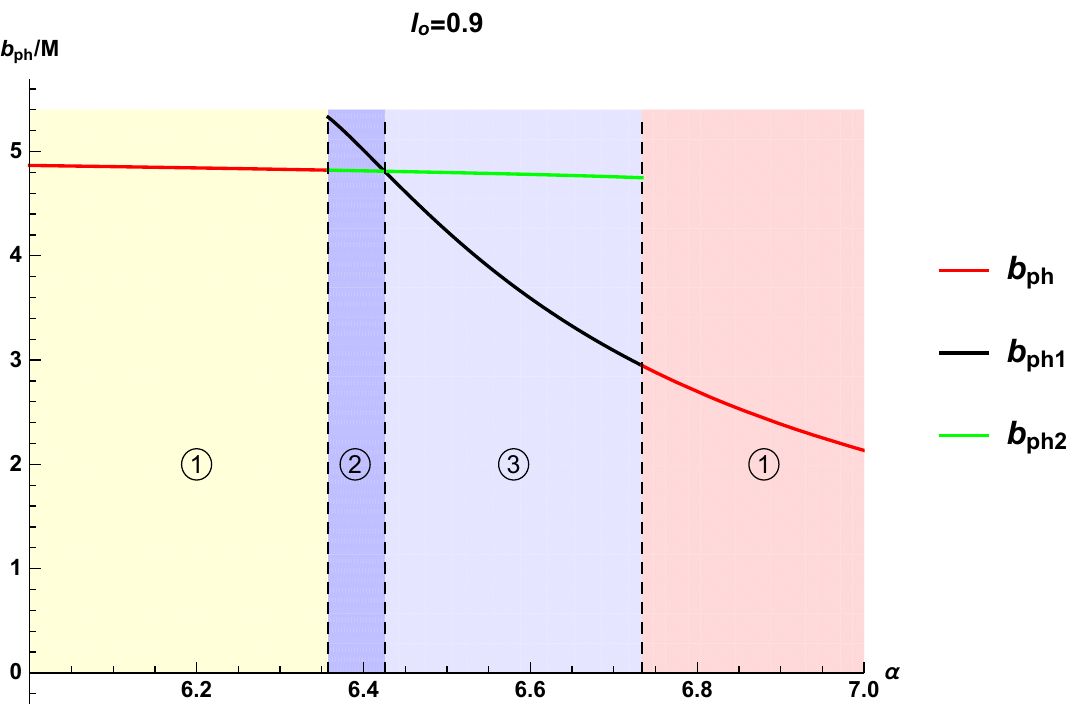}\label{fig:bc}}
   \caption{{\bf (a)}: The radius of photon sphere as a function of $\alpha$ with fixed $l_o=0.9$. In the yellow and pink regions, only one photon sphere exists while there exist two photon spheres in the blue region.  {\bf (b)}:  The critical impact parameter as a function of $\alpha$ with fixed $l_o=0.9$.  For configuration $\textcircled{1}$ ,  $b_{ph}$ which corresponds to the unique photon sphere also decreases as $\alpha$ increases.  In configuration $\textcircled{2}$, we have $b_{ph1}$ for the inner photon sphere is larger than $b_{ph2}$ for the outer photon sphere,  while  $b_{ph1}<b_{ph2}$ in configuration $\textcircled{3}$.}
\end{figure}

Here, with the help of \eqref{eq:formulabrph},  we shall explicitly show the photon spheres and its related critical impact parameters of the hairy Schwarzschild black hole. It is noted that in this sector, we have the quantities $(\alpha,l_o,r,b,M)$, which will be rescaled by $M$ to be dimensionless   $(\alpha,l_o/M,r/M,b/M,1)$. Therefore, from now on, all the physical quantities will be evaluated as the dimensionless ones, and so in the calculation we are safe to set $M=1$. For the sake of the discussion, we shall fix $l_o=0.9$ and tune $\alpha$ to include different configurations.  The radius of photon sphere(s) as a function of $\alpha$ is shown in FIG.\ref{fig:rp}. For most values of $\alpha$ (yellow and pink regions), only one photon sphere with radius $r_{ph}$ exists corresponding to the single maximum of the potential function, while their exists a region of $\alpha$ (blue region), in which two photon spheres appear corresponding to two {maximums} of the potential function, and we denotes the radius $r_{ph1}<r_{ph2}$.
It is obvious from FIG. \ref{fig:rp} that the radius of photon sphere decreases due to the additional fields in the current model. In fact, the photon spheres of compact objects with extra matters have been widely studied, but the extra matters could enhance the photon radius (see for examples \cite{Wang:2023vcv,Zeng:2021mok,Sheikhahmadi:2023jpb}) or decrease it (see for examples \cite{Pantig:2021zqe,He:2021htq,Guo:2021bwr}). So the effect of additional filed on the photon sphere is model dependent, and further studies are deserved for better understanding.

\begin{figure} [htbp]
\centering
\subfigure[\, typical behavior of potential for the configuration  $\textcircled{1}$.]
{\includegraphics[width=5cm]{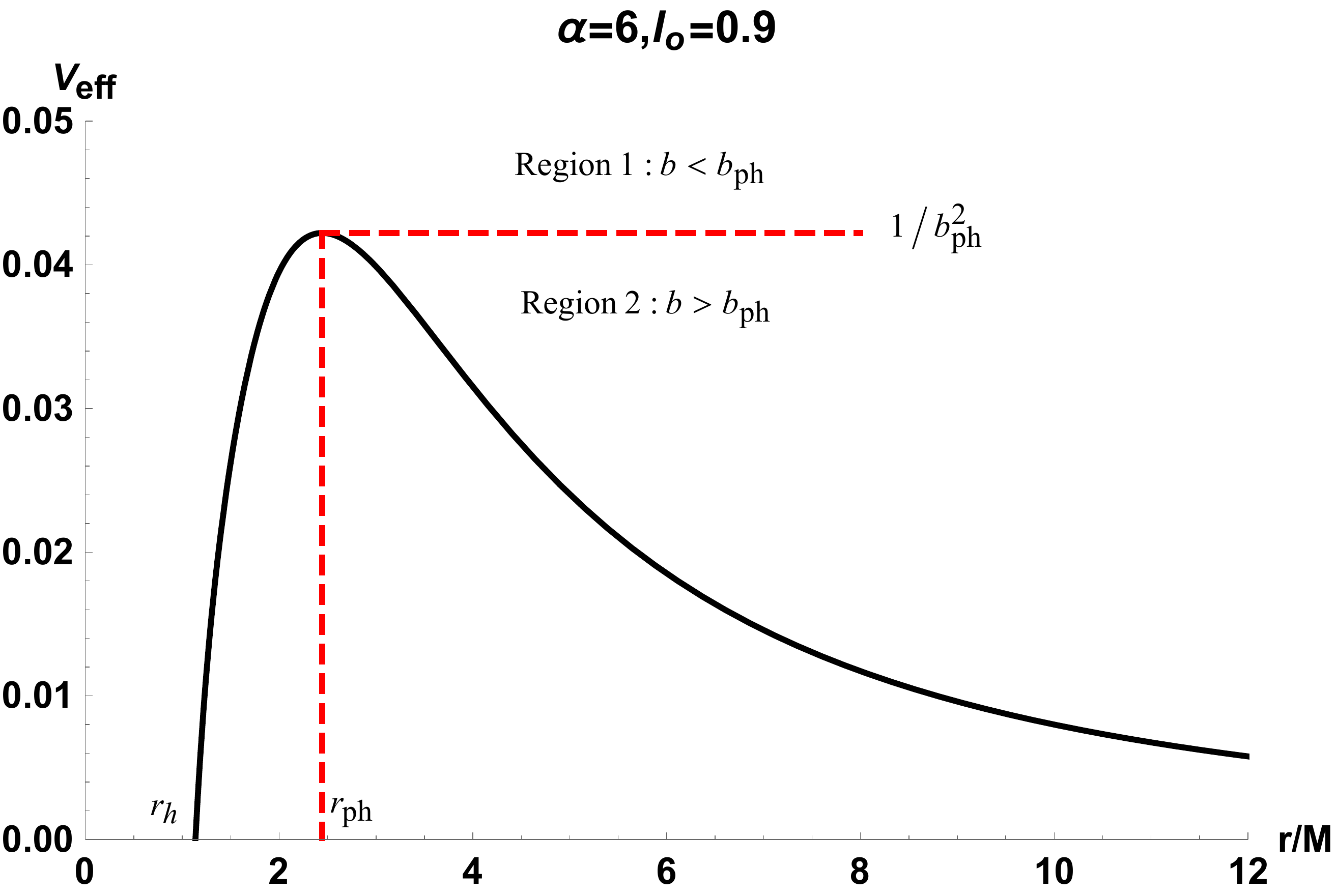} \label{fig:Veffa}}\hspace{1cm}
\subfigure[\, typical behavior of potential for the configuration  $\textcircled{2}$.]
{\includegraphics[width=5cm]{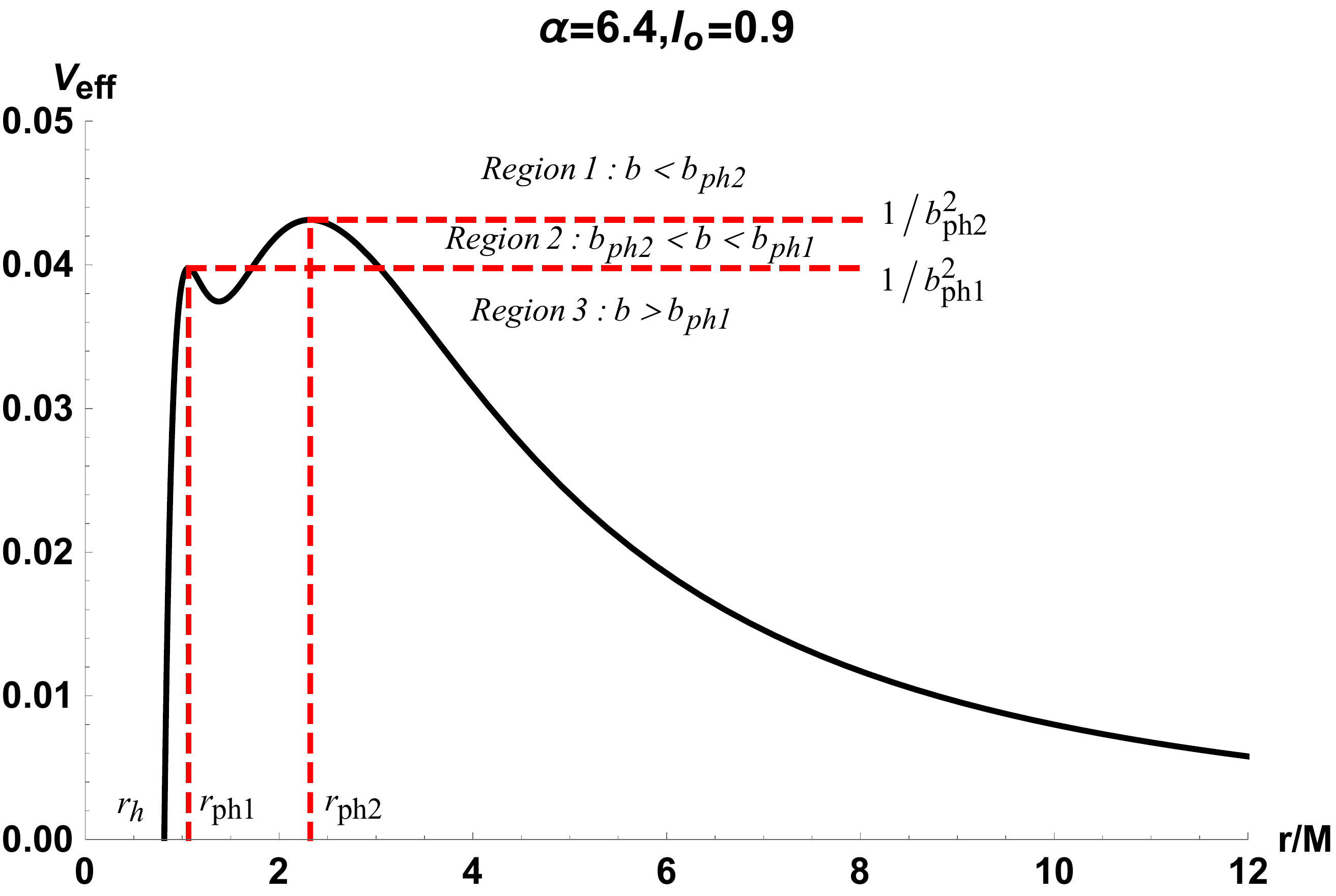} \label{fig:Veffb}}\hspace{1cm}
\subfigure[\, typical behavior of potential for the configuration  $\textcircled{3}$.]
{\includegraphics[width=5cm]{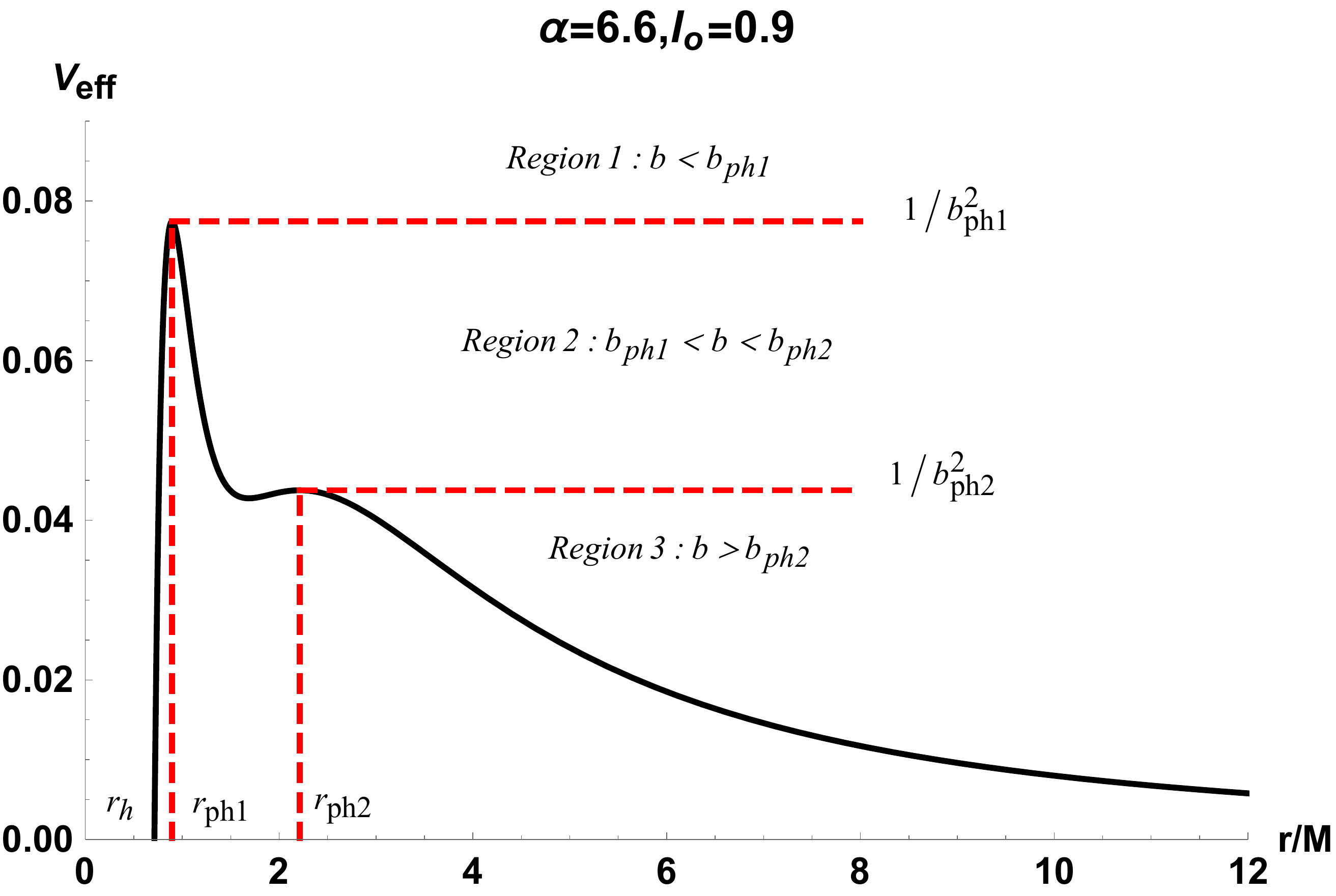} \label{fig:Veffc}}
\caption{The effective potential $V_{eff}$ as function of {radial coordinate} $r$ for three typical configurations depicted in FIG.\ref{fig:bc}. Here we have set $M=1$. Please see the main text for the understandings.}
\end{figure}

Then in FIG.\ref{fig:bc}, we depict the critical impact parameter(s). To explicitly distinguish the configurations, we only show the related results of $b_c$ for the  parameters in the central part of FIG.\ref{fig:rp}. Similarly, $b_{ph}$ also decreases as the $\alpha$ increases. In particular, in the case with double photon spheres, as $\alpha$ increases, $b_{ph1}$ for the inner photon sphere decreases faster and will then be smaller than $b_{ph2}$ for the outer photon sphere. From \eqref{eq:formulabrph}, this means that for some parameters, the maximum value of the potential function for inner photon sphere will become larger than that for outer photon sphere. Therefore, in terms of the behavior of $b_{ph}$, we shall classify the configurations into three types: $\textcircled{1}$ single $b_{ph}$ (yellow and pink regions), $\textcircled{2}$ $b_{ph1}>b_{ph2}$ (dark blue region) and  $\textcircled{3}$ $b_{ph1}<b_{ph2}$ (light blue region). The typical behavior of the potentials for the three types of configurations is presented in FIG.\ref{fig:Veffa}-Fig.\ref{fig:Veffc}.
%%%%%%%%%%%%%%%%%%%%%
%%%%%%%%%%%%%%%%%%%%%%%
\begin{itemize}
\item For the  configuration $\textcircled{1}$ shown in FIG.\ref{fig:Veffa}, as the photon moves closer to the black hole, the effective potential  first increases to the maximum value at $r_{ph}$ which corresponds to $b_{ph}$, and then decreases. For the photon with $b<b_{ph}$, it will be bounded by the effective potential and falls into the black hole. On the contrary, for the photon with $b>b_{ph}$, it will spread to infinity.
\item For the double photon spheres case, the configuration $\textcircled{2}$ with $b_{ph1}>b_{ph2}$ means that the maximum of the effective potential for the inner photon sphere is smaller than that for the outer photon sphere as shown in Fig.\ref{fig:Veffb}. Thus, the inner photon sphere indeed cannot escape from the binding of gravity of black hole to infinity. This implies that the fate of photon with the configuration $\textcircled{2}$ is in essence similar to that with configuration $\textcircled{1}$.
\item
Fig. \ref{fig:Veffc} shows the configuration $\textcircled{3}$ with the effective potential of the inner photon sphere larger than that of the outer photon sphere.  It is obvious that in this case  the light rays of the inner photon sphere may escape away from the binding of the black hole and reach the observer at infinity.
\end{itemize}
Moreover, in all cases,  the effective potential vanishes at the event horizon as expected. The above analysis on the fate of various photons will be checked by solving the geodesic equations with the use of ray tracing method in next section.

\section{Ray tracing and photon trajectory} \label{sec:photon trajectory}
We will solve the geodesic equations and figure out the photon trajectory around
the hairy Schwarzschild black hole, which is to pave the way to study the optical appearance of the black hole.
The trajectory of light ray is determined by the deformed geodesic equation
\begin{eqnarray}
\frac{dr}{d\phi}=\frac{\dot{r}}{\dot{\phi}}=\pm\frac{1}{r^2}\sqrt{\frac{1}{b^2}-V_{\text{eff}}(r)}.
\label{trajectory-light-ray}
\end{eqnarray}
To proceed, we use the backtracked ray-tracing method. In this method, the photon trajectory arriving to the observer's screen is backtracked by employing the above equation  to determine the point of the sky where it emits.  Thus, for the photon with impact parameter $b>b_{ph}$, it may turn around the central hairy black hole certain times, subsequently, we can divide the impact parameter region in terms of the total number of photon orbits $n=\frac{\phi}{2\pi}$
which  accounts the number of its intersection with the equatorial plane.
In details,  for the observer located at the north pole, the photons are classified into three classes: the first class is defined as the direct emission with $1/4<n\leq3/4$, where the light ray intersects the equatorial plane only once ($m=1$). The second class with $3/4<n\leq5/4$, where light ray intersects  the equatorial plane twice ($m=2$), corresponds to the lensed ring emission. The final class is photon ring emission where the light ray with $n>5/4$ crosses the accretion disk at least three times ($m=3$).

Then, we shall explore the total number of orbit as a function of impact parameter, and then show the photon trajectory seen by the aforementioned observer.  We intend to study the effect of the hairy parameters on the classification of the photon emission, and also check the difference between the  cases with single photon sphere and double photon spheres.

\subsection{Single photon sphere}
\begin{figure}[htbp]
\centering
      \includegraphics[width=6cm]{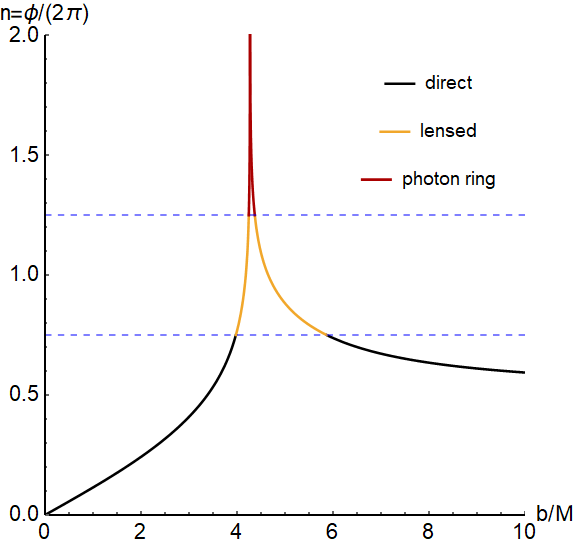} \hspace{1cm}
      \includegraphics[width=6cm]{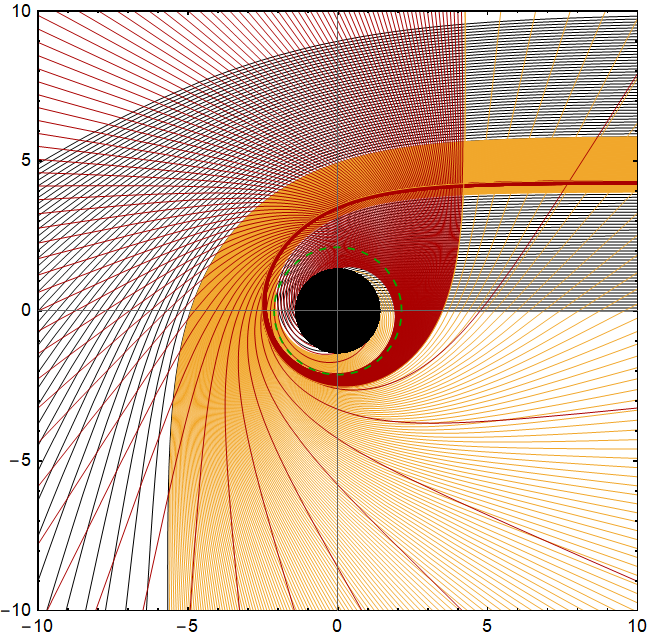}
\caption{{\bf Left}: the number of photon orbits $n$  as a function of the impact parameter $b$ for the hairy Schwarzschild black holes with $\alpha=2$ and $l_o=0.2$. {\bf Right}: a selection of associated photon trajectories in the Euclidean polar coordinates $(r,\phi)$. The black curves, gold curves, and red curves correspond to the direct emissions ($n<3/4$), lensed ring emissions ($3/4<n<5/4$), and photon ring emissions ($n>5/4$), respectively. The black disk and the green dashed curves denote the event horizon and photon sphere.}
\label{fig:orbits-rings-a-2}
\end{figure}

For the configuration with single photon sphere, the typical results of the total number of photon orbits and the photon trajectory in the observer's sky are shown in Fig.\ref{fig:orbits-rings-a-2} where we choose $\alpha=2$ and $l_o=0.2$ as a sample.  At $b=b_{ph}=4.27181$, the total number of orbits $n$ is divergent, and the photon travels around the black hole many times. For $b<b_{ph}$, as  $b$ increases,  the total number $n$ of orbits increases, while for  $b>b_{ph}$, $n$  decreases as $b$ increases. The red, gold, and black curves in the left plot correspond to the photon ring emissions with $b\in(4.24664,4.37160)$ , lensed ring emissions with $b \in(3.97754,4.24664)\cup (4.37160,5.87177)$, and direct emissions with $b\notin(3.97754,5.87177)$, which are denoted by the trajectories with the same colors in the right plot where their intersections with  the equatorial plane are clear.

To study the effects of hairy parameters on this phenomena, we do the parallel calculations for various values of parameters. The data of event horizon, photon sphere radius, critical impact parameters and regions for direct emissions, lensed ring emissions and photon ring emissions for fixed $\alpha=2$ with samples of  $l_o$, are listed in Table \ref{table01}. We see that as the primary hair parameter $l_o$ increases, the critical impact parameter $b_{ph}$ increases, and all border values of $b$ for direct, lensed ring and photon ring emissions gradually increase. But the width of photon ring and lensed ring emissions decrease, which implies that the stronger hairy charge of the black hole may make the rings more difficult to be detected.

%%%%%%%%%%%%%
\begin{table}[htbp]
{\centering
\begin{tabular}{|c|c|c|c|c|c|c|c|}
  \hline
  $\alpha$ & $l_o$& $r_h$& $r_{ph}$& $b_{ph}$&Direct emission&Lensed ring emission& Photon ring emission\\
  \hline
  2 &0.2 &1.41177 &2.12643 &4.27181 &$b<3.97754$; $b>5.87177$ &$3.97754<b<4.24664$; $4.37160<b<5.87177$ &$4.24664<b<4.37160$\\
  \hline
  2 &0.4 &1.55466 &2.40392 &4.65128 &$b<4.34062$; $b>6.01986$ &$4.34062<b<4.62465$; $4.73567<b<6.01986$ &$4.62465<b<4.73567$\\
  \hline
  2 &0.6 &1.70033 &2.65288 &4.91526 &$b<4.63092$; $b>6.10425$ &$4.63092<b<4.89398$; $4.97751<b<6.10425$ &$4.89398<b<4.97751$\\
  \hline
 \multicolumn{2}{|c|}{GR}  &2       &3       &5.19615 &$b<5.01514$; $b>6.16757$ &$5.01514<b<5.18781$; $5.22794<b<6.16757$ &$5.18781<b<5.22794$\\
  \hline
\end{tabular}
\caption{The event horizon, photon sphere radius, critical impact parameters of hairy Schwarzschild black holes with different parameter $l_o$. The ranges of impact parameter $b$ correspond to the direct, lensed ring, and photon ring emissions of the hairy Schwarzschild black holes for different $l_o$. Here, we fix $M=1$.
\label{table01} }}
\end{table}
%%%%%%%%

Moreover, those results for fixed $l_o=0.2$ with different derivation $\alpha$  are listed in Table \ref{table02}. We see that in contract to the effect of $l_0$, larger $\alpha$ corresponds to smaller event horizon, photon sphere and various impact parameter borders of emissions. Moreover,
with the increasing of $\alpha$, the width of the photon ring emissions and the lensed rings emissions in the region $b<b_{ph}$ first increases and then decreases, while the width of the lensed ring emissions with $b>b_{ph}$ increases, indicating that the light deflection to the black hole becomes more intense for larger $\alpha$.
%%%%%%%%%%%%%%%%%%%%%%
\begin{table}
{\centering
\begin{tabular}{|c|c|c|c|c|c|c|c|}
  \hline
  $\alpha$ & $l_o$& $r_h$& $r_{ph}$& $b_{ph}$&Direct emission&Lensed ring emission& Photon ring emission\\
  \hline
  \multicolumn{2}{|c|}{GR}   &2       &3       &5.19615 &$b<5.01514$; $b>6.16757$ &$5.01514<b<5.18781$; $5.22794<b<6.16757$ &$5.18781<b<5.22794$\\
  \hline
  1 &0.2 &1.74958 &2.65630 &4.85508 &$b<4.62164$; $b>6.03597$ &$4.62164<b<4.84057$; $4.90664<b<6.03597$ &$4.84057<b<4.90665$\\
  \hline
  2 &0.2 &1.41177 &2.12643 &4.27180 &$b<3.97754$; $b>5.87177$ &$3.97754<b<4.24664$; $4.37160<b<5.87177$ &$4.24664<b<4.37160$\\
  \hline
  3 &0.2 &1.01406 &1.42810 &3.09224 &$b<2.89496$; $b>5.64942$ &$2.89496<b<3.07932$; $3.18976<b<5.64942$ &$3.07932<b<3.18976$\\
  \hline
  3.5&0.2&0.84356 &1.16011 &2.36596 &$b<2.27206$; $b>5.49642$ &$2.27206<b<2.36240$; $2.39589<b<5.49642$ &$2.36240<b<2.39589$\\
  \hline
  4 &0.2 &0.70974 &0.96860 &1.77229 &$b<1.73704$; $b>5.28260$ &$1.73704<b<1.77156$; $1.77830<b<5.28260$ &$1.77156<b<1.77830$\\
  \hline
\end{tabular}
\caption{The event horizon, photon sphere radius, critical impact parameters of hairy Schwarzschild black holes with different parameter $\alpha$. The ranges of impact parameter $b$ correspond to the direct, lensed ring, and photon ring emissions of the hairy Schwarzschild black holes for different $\alpha$. Here, we fix $M=1$.
\label{table02} }}
\end{table}

\subsection{Double photon spheres}
\begin{figure}[htbp]
\centering
\subfigure[\, $\alpha=6.4,l_o=0.9$]
{\includegraphics[width=6cm]{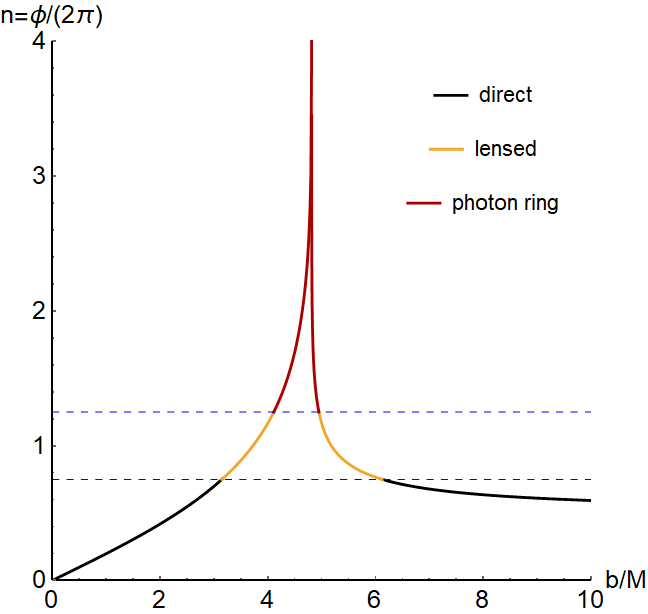}\hspace{1cm}
      \includegraphics[width=6cm]{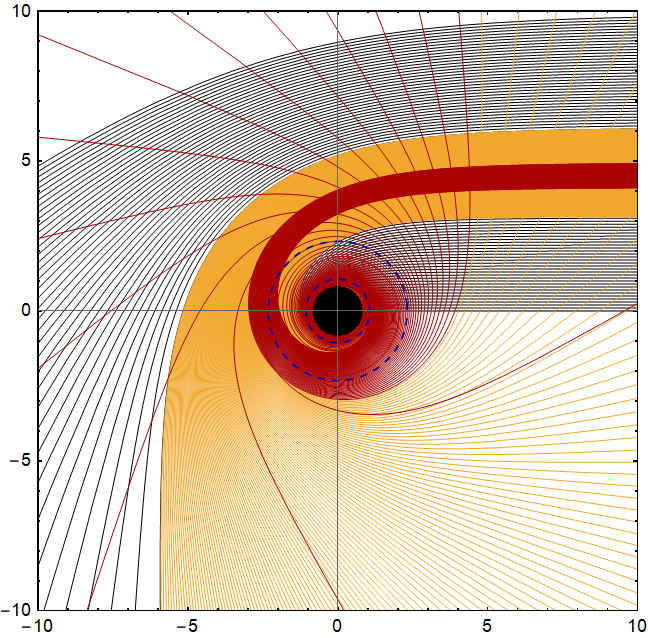}
 \label{fig:orbits-rings-double-peak1}
}
\subfigure[\, $\alpha=6.6,l_o=0.9$]
{\includegraphics[width=6cm]{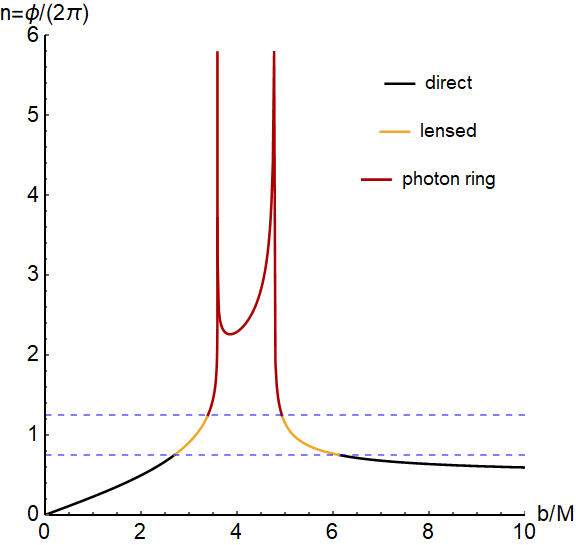}\hspace{1cm}
      \includegraphics[width=6cm]{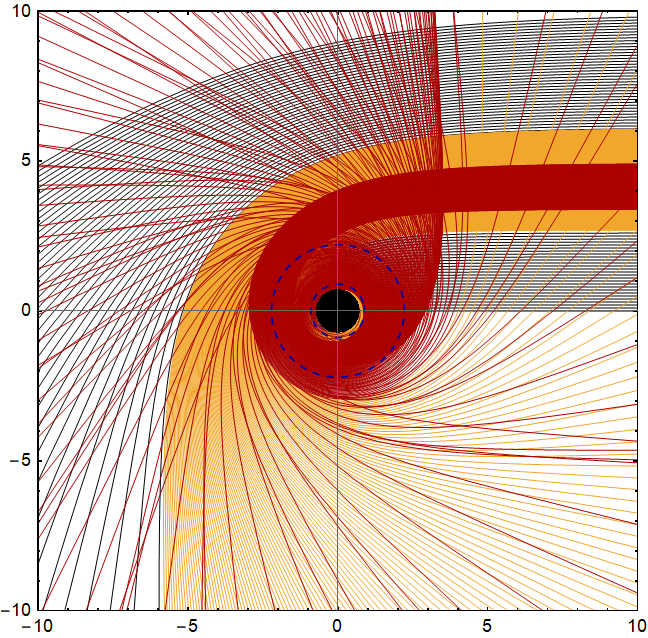}
\label{fig:orbits-rings-double-peak2}
}
\caption{{\bf Left}: the number of photon orbits $n$  as a function of the impact parameter $b$ for the hairy Schwarzschild black holes. {\bf Right}: a selection of associated photon trajectories in the Euclidean polar coordinates $(r,\phi)$. The black curves, gold curves, and red curves correspond to the direct emissions ($n<3/4$), lensed ring emissions ($3/4<n<5/4$), and photon ring emissions ($n>5/4$), respectively. The black disk and the green dashed curves denote the event horizon and photon sphere. The upper panel is for $\alpha=6.4$ and $l_o=0.9$ while the bottom panel is for $\alpha=6.6$ and $l_o=0.9$.}
\label{fig:orbits-rings-double-peak}
\end{figure}

We move on to discuss the direct emissions, lensed ring emissions and photon ring emissions of hairy Schwarzschild black holes with double photon spheres.  The total number of photon orbits and the photon trajectory for the configuration $\textcircled{2}$ with $\alpha=6.4$ and $l_o=0.9$ are shown in FIG.\ref{fig:orbits-rings-double-peak1}, in which the behaviors are similar to those in FIG.\ref{fig:orbits-rings-a-2} , only with wider photon ring emissions and lensed ring emissions. {This is indeed expected because as we aforementioned in last section, that the double photon spheres with  configuration $\textcircled{2}$ is in essence similar to the single photon sphere, and the inner photon sphere cannot escape to the distance observer. Thus, only one peak is observed in the total number of photon orbit, which is located at $b=b_{ph2}$, while the inner photon sphere is bounded by the black hole and cannot be captured by the observer.
Meanwhile, we show the results for the configuration $\textcircled{3}$ with $\alpha=6.6$ and $l_o=0.9$ in FIG. \ref{fig:orbits-rings-double-peak2}.
Obviously, the behavior of $n$ is significantly different from that for either single photon sphere or double photon spheres with  configuration $\textcircled{2}$. Two peaks in $n=\phi/2\pi$ are observed, which corresponds to two photon spheres captured by the observer at the related  critical impact parameters.} The photon ring emission in this case is significantly enhanced, which will be more pronounced when the further apart the  two critical impact parameters are. In addition, the ranges of direct, lensed ring, and photon ring emissions  with respect to the impact parameter $b$ are solved and the results are listed in Table \ref{table03}.

%%%%%%%%%%%%%
\begin{table}
{\centering
\begin{tabular}{|c|c|c|c|c|c|c|c|c|c|}
  \hline
  $\alpha$ &$l_o$ &$r_h$ &$r_{ph1}$ &$r_{ph2}$ &$b_{ph1}$ &$b_{ph2}$ &Direct emission &Lensed ring emission &Photon ring emission\\
  \hline
  6.4 &0.9 &0.8143 &1.0659 &2.3209 &- &4.8148 &$b<3.1478$; $b>6.1321$ &$3.1478<b<4.1116$; $4.9497<b<6.1321$ &$4.1116<b<4.9497$\\
  \hline
  6.6 &0.9   &0.7127 &0.8954 &2.1237 &3.5935 &4.7803 &$b<2.6974$; $b>6.1310$ &$2.6974<b<3.3956$; $4.9354<b<6.1310$ &$3.3956<b<4.9354$\\
  \hline
\end{tabular}
\caption{In the case of double-photon sphere, the event horizon, photon sphere radius, critical impact parameters of hairy Schwarzschild black holes. The ranges of impact parameter $b$ correspond to the direct, lensed ring, and photon ring emissions of the hairy Schwarzschild black holes with double photon spheres. Here, we fix $M=1$.
\label{table03} }}
\end{table}
%%%%%%%%%%%%%%%%%%%

According to the above analysis, in view of the observer in the north pole, the light rays around the black hole with two photon spheres with configuration \textcircled{2} are  essentially the same as that with single photon sphere. Only  the double photon spheres with configuration $\textcircled{3}$ will bring additional observation features, as also addressed in \cite{Gan:2021xdl,Guo:2022muy,Chen:2022qrw}. So in the following study on optical appearance, we will only consider the cases with single photon sphere and double photon spheres with configuration $\textcircled{3}$, which for convenience will be denoted by single photon sphere and double photon spheres, respectively.

So far, the analysis concentrates on the central depression of the image of the hairy Schwarzschild black hole seen by a far-away observer and the shapes of the light rings, which are idealized observables. Considering that the realistic astrophysical images mainly consist of the physics
of the accretions around the central objects, it is of interest to explore the images of hairy Schwarzschild black hole illuminated by accretions.  For convenience, we will focus on static and thin (geometrically and optically) accretions, which are disk and spherically symmetric, respectively.
Besides the effects of hairy parameters on the optical appearances,  we are especially interested in differentiating the optical appearances of the hairy Schwarzschild  black hole with double photon spheres from that with single photon sphere. To this end,  we choose three couples of hairy parameters, i.e, $\alpha=2~\&~l_o=0.2$,  $\alpha=2~\&~l_o=0.6$  and $\alpha=3 ~ \&~ l_o=0.2$ to discuss the optical properties of hairy Schwarzschild black hole with single photon sphere, while $\alpha=6.6~\&~l_o=0.9$ for double photon spheres.

\section{Shadows and rings with static thin accretion disk}\label{sec:thin disk accretions}
In this section, we will explore the images of hairy Schwarzschild black hole illuminated by the optically and geometrically thin accretion disk, which is located at rest on the equatorial plane around the black hole, viewed face-on.
Since the light ray will extract energy from the thin accretion disk each time when passing through it, so different types of emissions will contribute differently to the observed light intensity. The analysis in previous section indicates that the  hair has a significant effect on  widths of various emissions. So it is interesting to further study the observed intensities and see the hairy Schwarzschild black hole's observational appearance.

\subsection{Observed specific intensities and transfer functions}
Considering that the thin accretion disk emits isotropically in the rest frame of static worldlines, the specific intensity received by the observer with emission frequency $\nu_e$ is
\begin{equation}
I_{o}(r, \nu_o)=g^3 I_{e}(r,\nu_e),
\end{equation}
where $g=\nu_o/\nu_e=\sqrt{f(r)}$ is the redshift factor, and $I_{e}(r,\nu_e)$ is the specific intensity of the accretion disk.
%and $\nu_o$ is observed frequency.
The total observed intensity $I_{obs}(r)$ can be obtained by integrating all observed frequencies of $I_{o}(r, \nu_o)$ written as
\begin{equation}
I_{obs}(r)=\int I_{o}(r, \nu_o) d\nu_o=\int g^4 I_{e}(r,\nu_e) d\nu_e=f(r)^2 I_{em}(r),
\end{equation}
where we denote $I_{em}(r)=\int I_{e}(r,\nu_e) d\nu_e$ as the total emitted intensity. We note that if the trajectory of photon followed backward from the observer intersects the disk, the photon from accretion disk emission will contribute the brightness to the observer \cite{Gralla:2019xty}. Thus, ignoring the absorption, the total observed intensities are determined by each intersection, yielding
\begin{equation}
I_{obs}(b)=\sum_{m}f(r)^2I_{em}(r)|_{r=r_m (b)}, \label{eqtransfer}
\end{equation}
where $r_m(b)$ denotes for the coordinate of the $m-th$ intersecting position between the light ray emitted with impact parameter $b$ and the accretion disk. $r_m (b)$ is also known as the transfer function because it describes the mapping from the impact parameter of the photon to the $m-$th  hitting position on the disk, and its slope $dr/db$ describes the demagnification factor at each $b$ \cite{Gralla:2019xty}.

\begin{figure} [htbp]
\centering
\subfigure[\, $\alpha=2, l_o=0.2$.]
{\includegraphics[width=7cm]{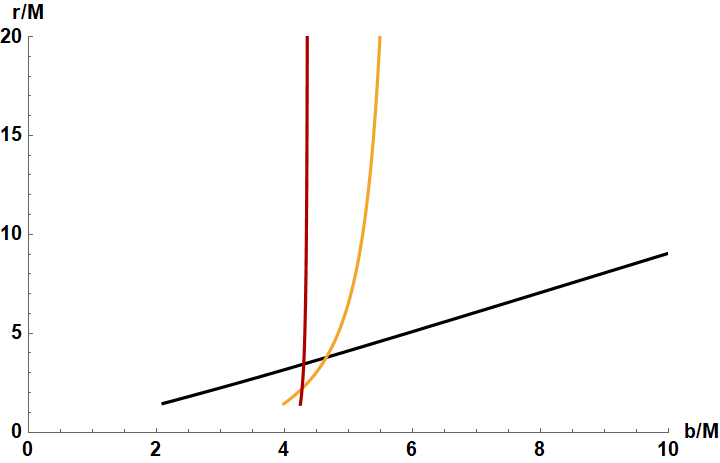}}\hspace{1cm}
\subfigure[\, $\alpha=6.6, l_o=0.9$.]
{\includegraphics[width=7cm]{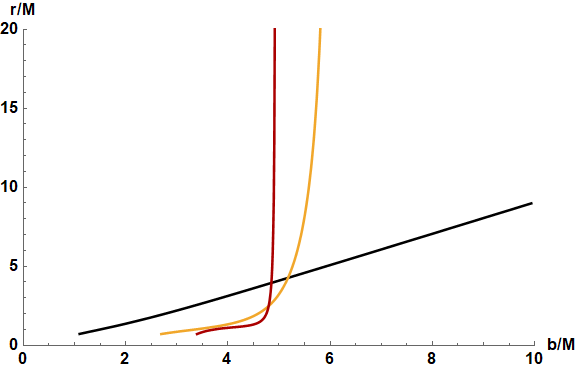}}
 \caption{The first three transfer functions of the hairy Schwarzschild black holes with single photon sphere (left) and   double photon spheres (right). They represent the radial
coordinate of the first (black), second (gold), and third (red) intersections with the emission, respectively.  Here we have set $M=1$.\label{fig:transfer function}}
\end{figure}

So, before studying the total observed intensities, we have to evaluate the transfer functions.
We shall focus on the first three transfer functions since the higher cases contribute much less to the total luminosity. {As illustrated in \cite{Wang:2023vcv}}, the first transfer function  corresponds to the direct image originating from direct, lensed and photon rings emission; the second transfer function can origin from lensed ring and photon ring emission; while the third transfer function can only origin from photon ring emission. In FIG.\ref{fig:transfer function}, we depict the transfer functions for the hairy Schwarzschild black hole with  single photon sphere ($\alpha=2, l_o=0.2$) and that with double photon spheres ($\alpha=6.6, l_o=0.9$). We can read off the following properties. (i) In both cases, the slope of the first transfer function is almost $1$, which means that this direct image can be seen as the source profile after redshift. (ii) For the hairy Schwarzschild black hole with single photon sphere, the slopes of the second and third transfer functions are much more than that of the first one, and $r_3(b)$ is steeper than $r_2(b)$.  This implies that in this case the first transfer function will give the largest contribution to the total luminosity and  the second and third ones are  highly demagnified. Additionally, we also check the effects of hairy parameters on the transfer functions. They have some effects on the width of the second and three transfer functions and their slopes, but comparing to the contribution from the first transfer function, those from others are still insignificant.
(iii) For the hairy Schwarzschild black hole with double photon spheres,  the widths of second and third transfer functions have been significantly widened. In particular, for $b\lesssim4.6$ their slopes are even smaller than that for first transfer function, indicating that they are not  demagnified. These imply that due to the existence of double photon spheres,  the second and third transfer functions could make important contributions to  the total luminosity.

With the transfer function in hands, Eq.\eqref{eqtransfer} indicates that we can then evaluate the observed  intensities from each
transfer function and so the total observed intensity, once the emission function is given.  It is natural to  regard the brightness contributed from the first, second and third transfer function as the direct, lensed ring and photon ring intensity, respectively. Next, to testify the  key  properties  read from the transfer functions,  we will consider some specific emission profiles of accretion disk, and then evaluate the total observed  intensity in terms of each kind of intensity.

\subsection{Optical appearances}
We shall consider the following two toy-models emission functions \cite{Wang:2022yvi,Yang:2022btw} to figure out the optical appearances of the hairy Schwarzschild black hole.

\subsubsection{Model I}

In Model I, we consider that the emission of the accretion disk starts from the innermost {stable} circular orbit $r_{isco}$, and the emission specific intensity is attenuated by the second-order function of the radial coordinate
%%%%%%%%%%%%%%%%%%
\begin{figure}[H]
\begin{minipage}{0.5\linewidth}
\begin{eqnarray}
        I_{em}^{I}(r)=
        \begin{cases}
            I_o\frac{1}{[r-(r_{isco}-1)]^2}                 ~~~~~~~& r>r_{isco}\\
            \\
            0,                                            ~~~~~~~& r\leq r_{isco}
        \end{cases} ,  ~~
    \label{obser1}
\end{eqnarray}
\end{minipage}
\hfill
\begin{minipage}{0.5\linewidth}
 \centerline{\includegraphics[width=1.5in]{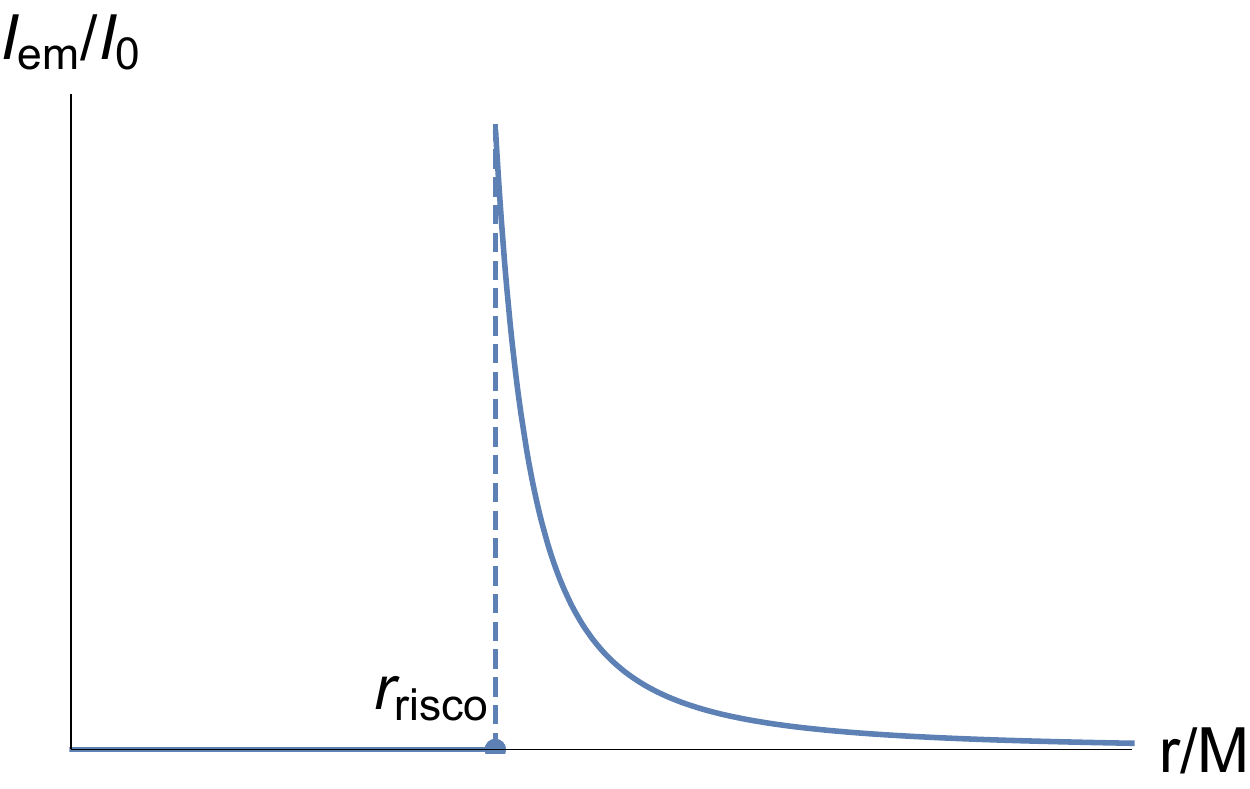}}
\end{minipage}
\end{figure}
Here $I_0$ is the maximum intensity (the same below) and the $r_{isco}$ of the hairy Schwarzschild (Kerr) black hole was calculated by some of us in \cite{Wu:2023wld}. The observed intensities and  images of the hairy Schwarzschild black hole illuminated by the above thin accretion model I are shown in FIG. \ref{fig:single-apperanceM1} (single photon sphere) and FIG. \ref{fig:double-apperanceM1} (double photon spheres) . In the figures,  the leftmost column shows the different observed intensities originated from the direct (black), lensed ring (gold) and photon ring (red) intensity respectively; then we present the total observed intensities which are translated  into the  optical appearances in their right sides. The same layout will also be used in the other model.
%%%%%%%%%
\begin{figure}[htbp]
\centering
\subfigure[\, $\alpha=2,l_o=0.2$]
{
\includegraphics[width=4cm]{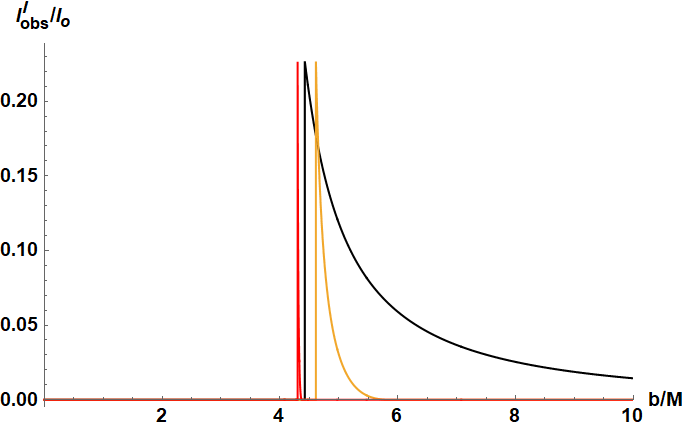}
      \includegraphics[width=4cm]{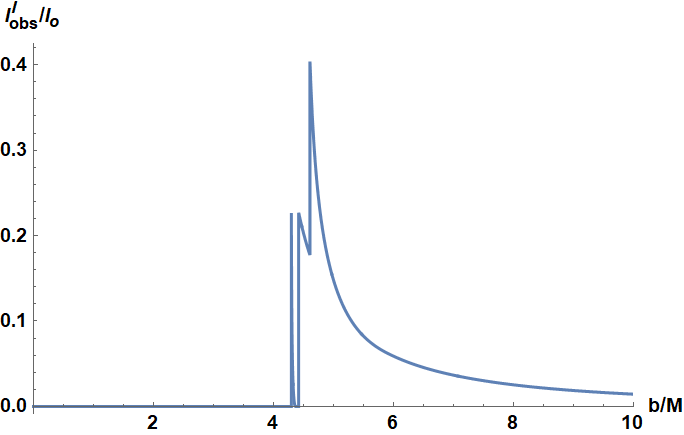}
      \includegraphics[width=4cm]{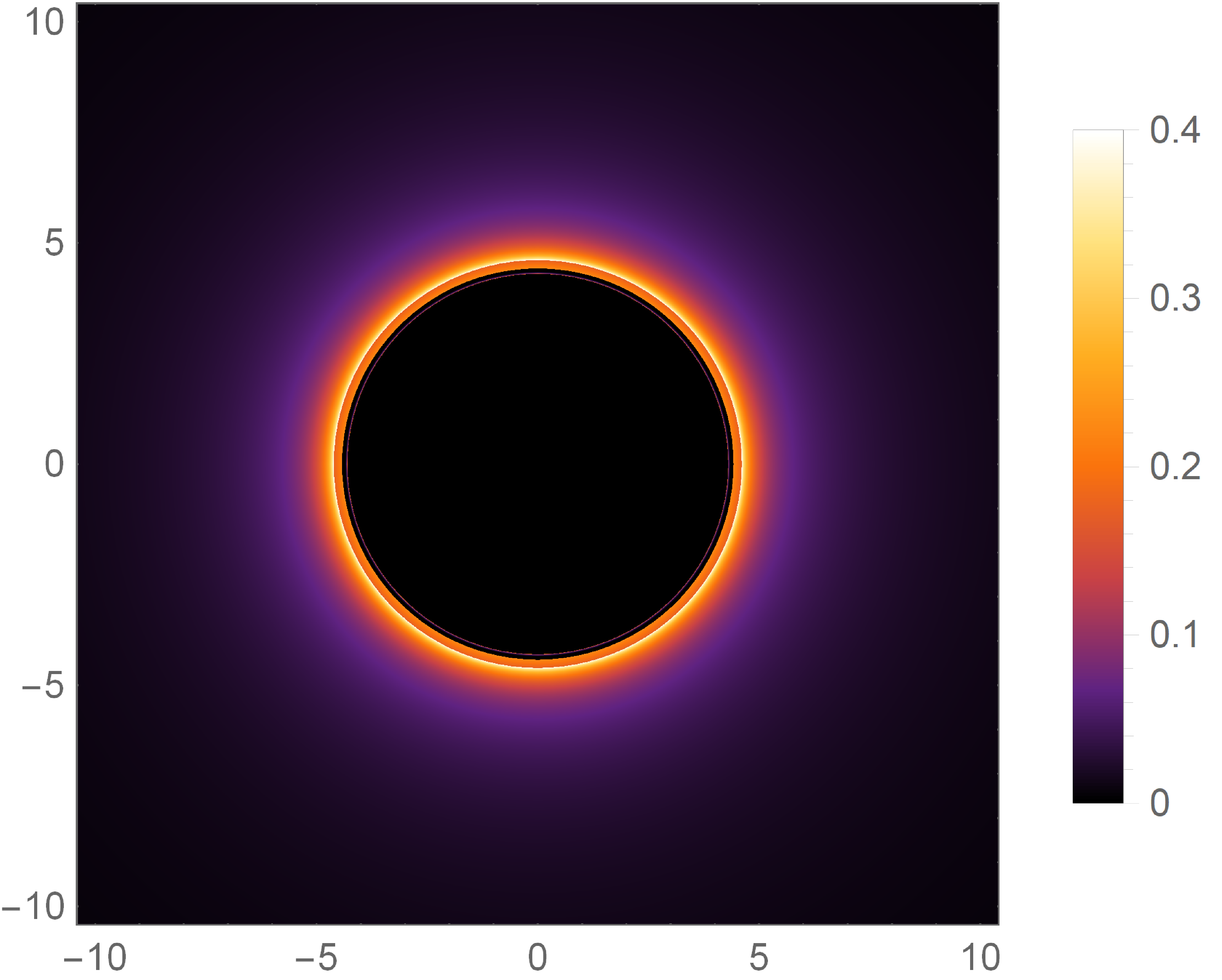}
      \includegraphics[width=3.2cm]{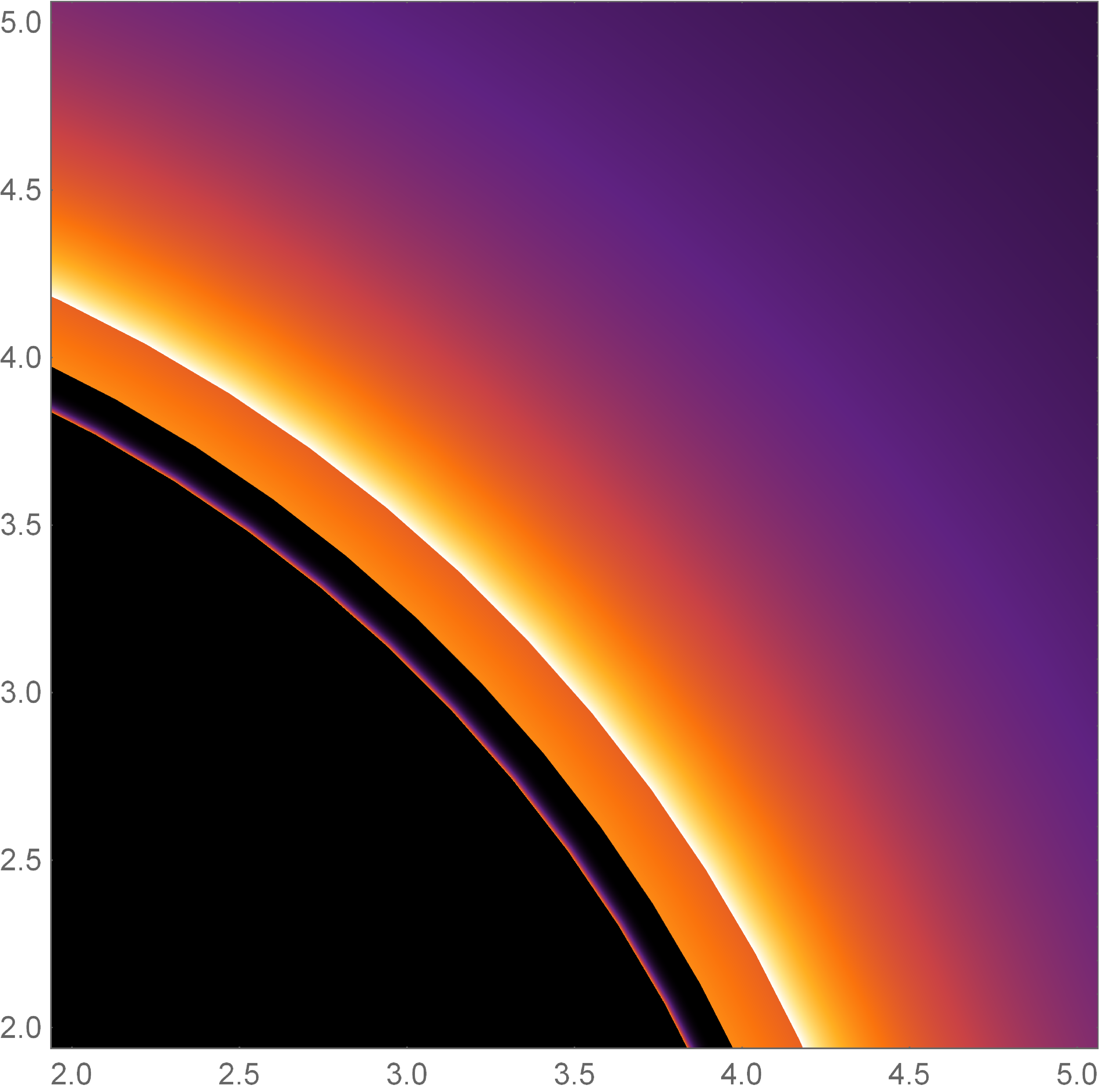}
\label{fig:single-apperanceM1-1}
}
\subfigure[\, $\alpha=2,l_o=0.6$]
{
\includegraphics[width=4cm]{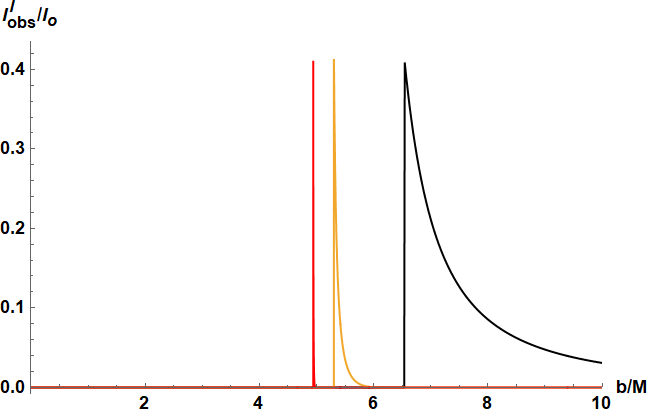}
\includegraphics[width=4cm]{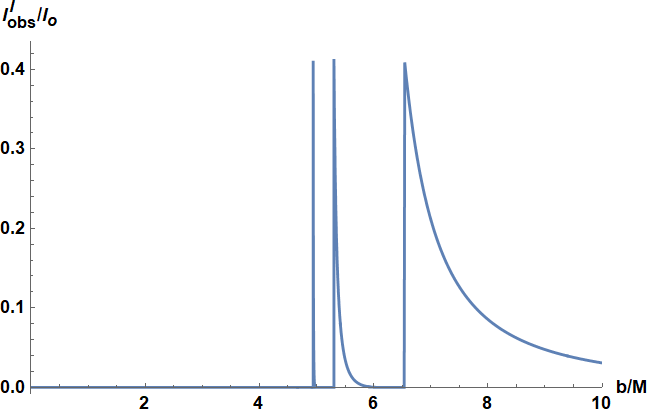}
\includegraphics[width=4cm]{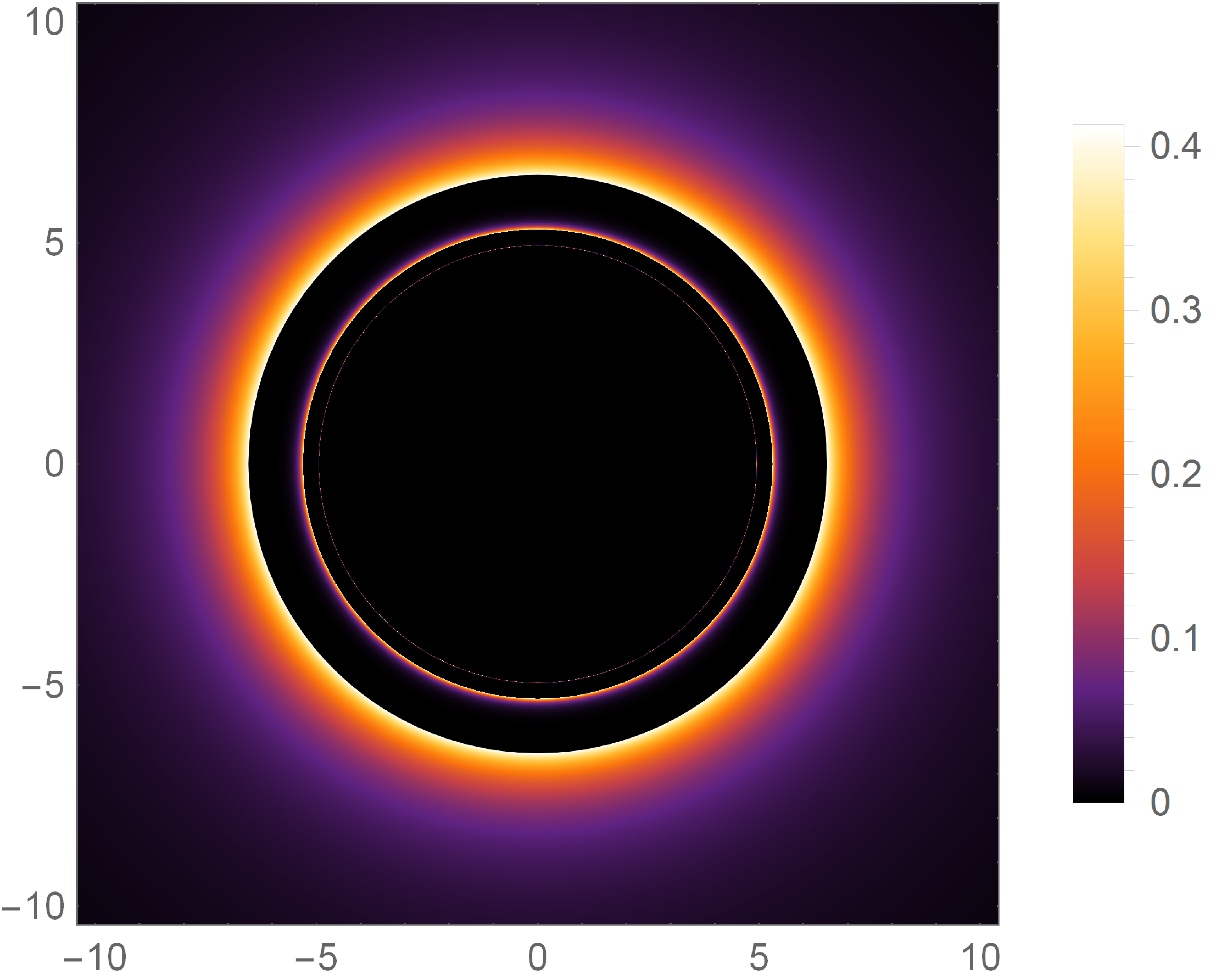}
\includegraphics[width=3.2cm]{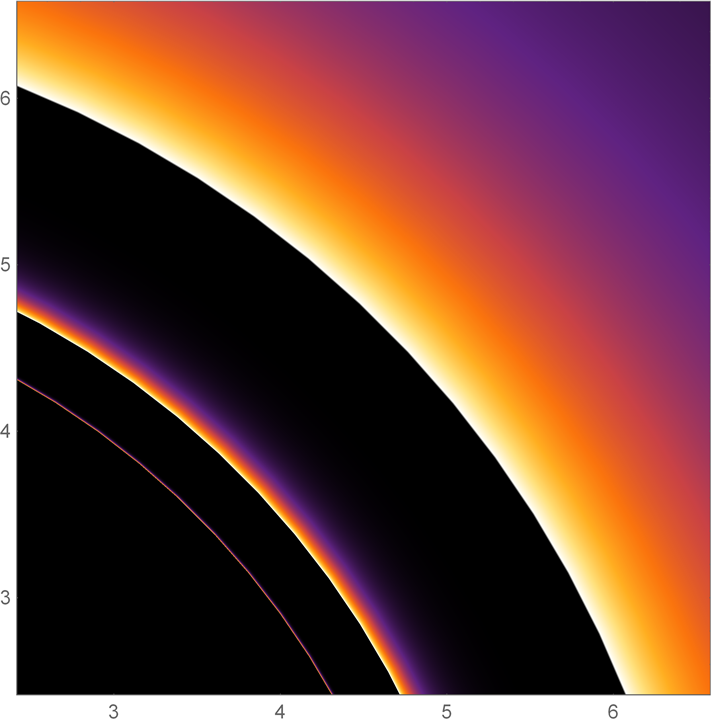}
\label{fig:single-apperanceM1-2}
}
\subfigure[\, $\alpha=3,l_o=0.2$]
{
\includegraphics[width=4cm]{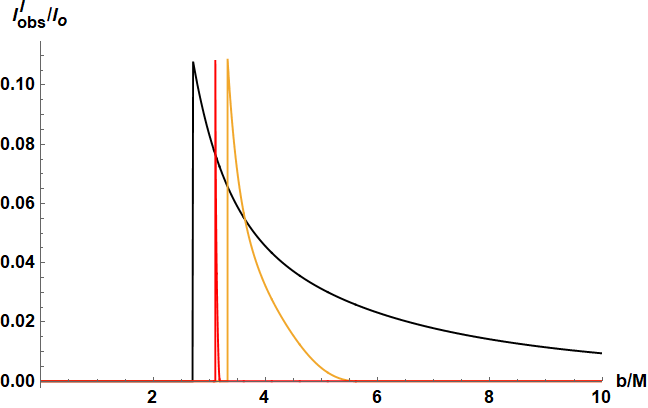}
\includegraphics[width=4cm]{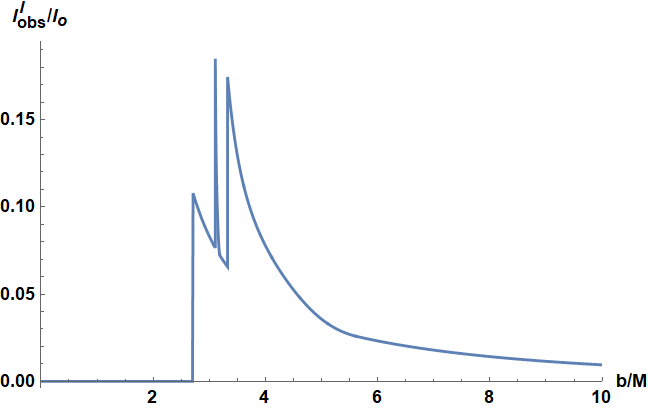}
\includegraphics[width=4cm]{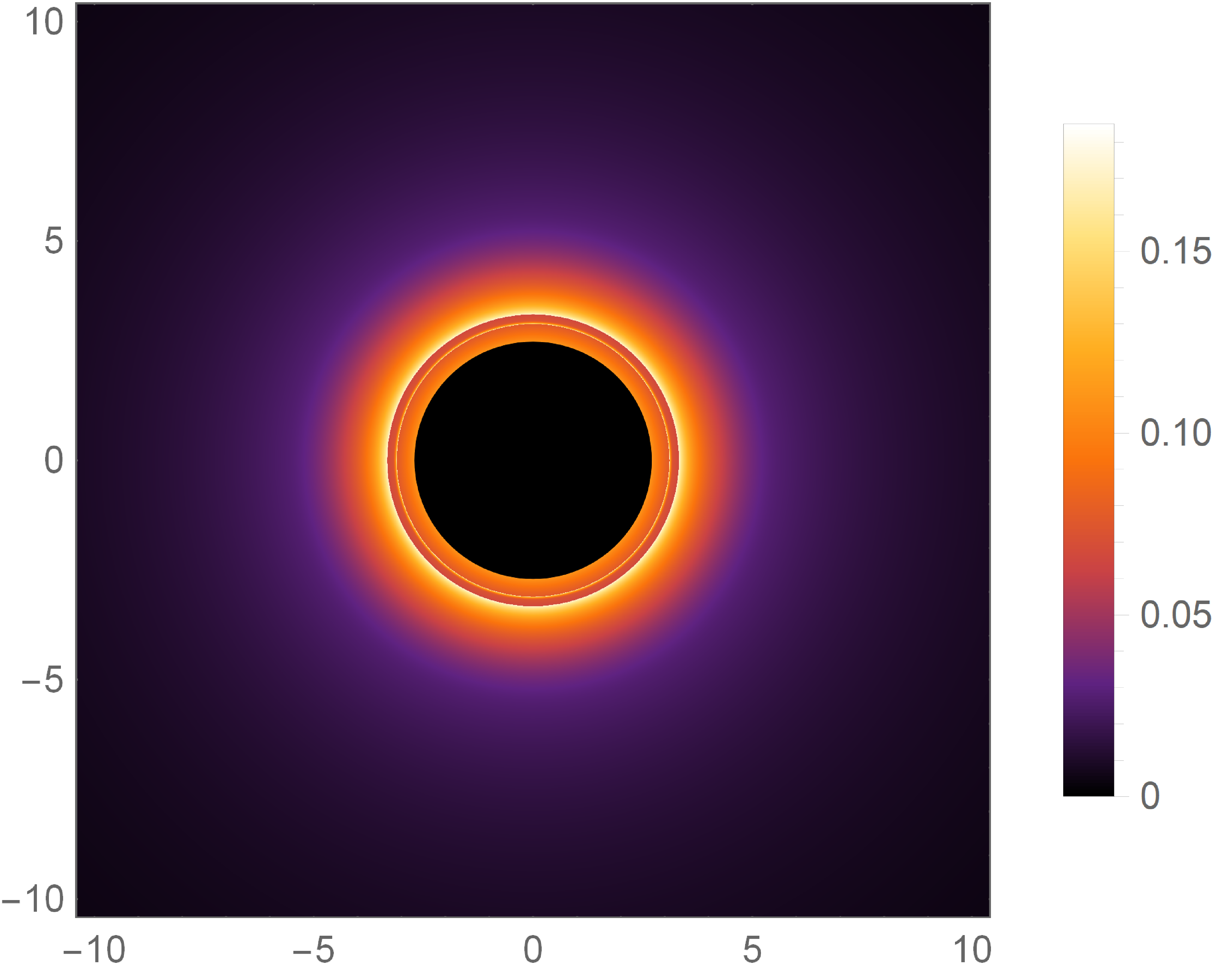}
\includegraphics[width=3.2cm]{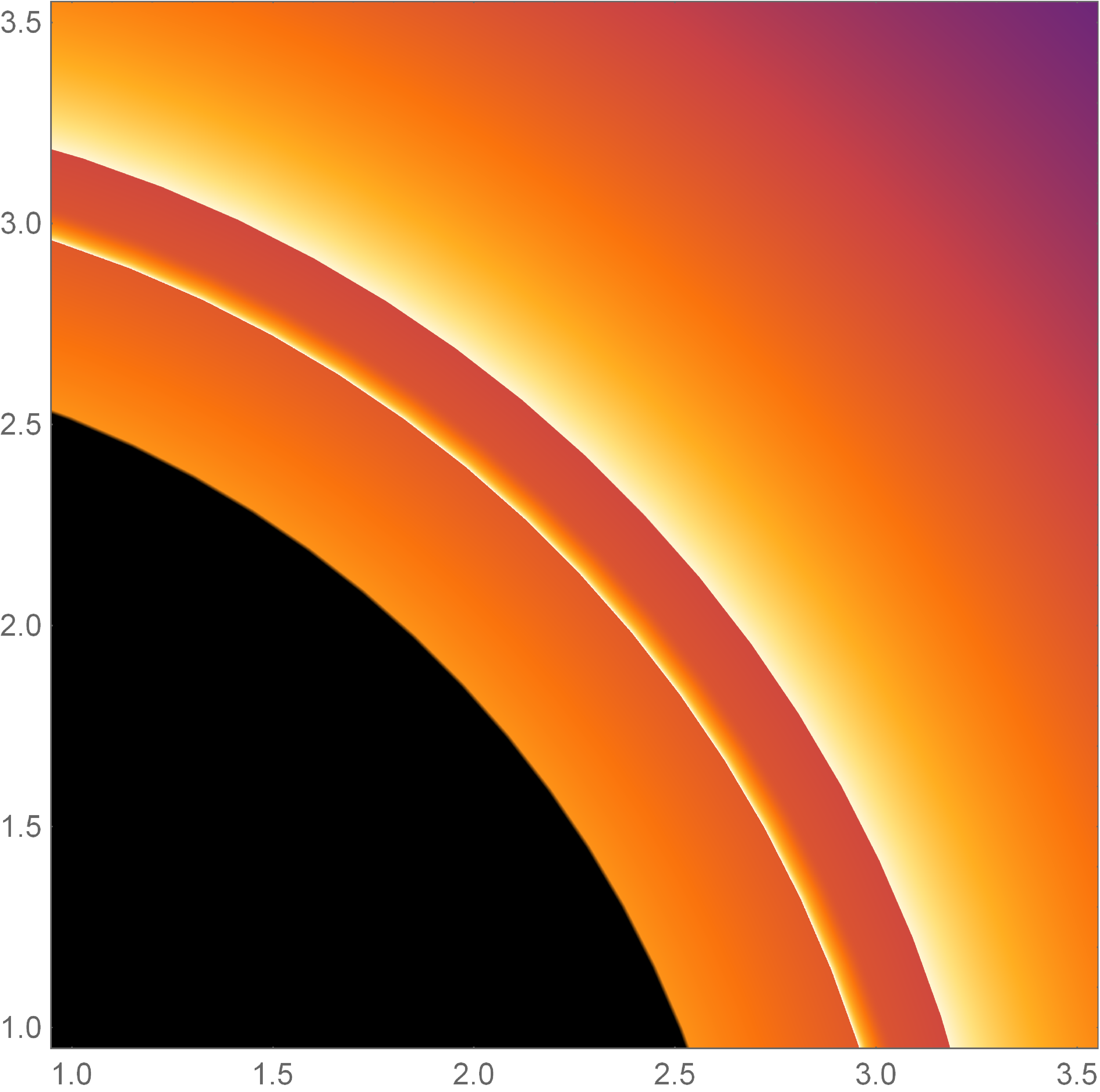}
 \label{fig:single-apperanceM1-3}
}
\caption{Observational appearances of the Model I \eqref{obser1} of a thin disk for different $\alpha$ and $l_o$ with $M=1$. \textbf{First column:} the different observed intensities originated from the first (black), second (gold) and third (red) transfer function in Eq. \eqref{eqtransfer} respectively. \textbf{Second column:} the total observed intensities $I_{obs}/I_{o}$ as a function of impact parameter $b$. \textbf{Third column:} the optical appearance: the distribution of observed intensities into two-dimensional disks. \textbf{Fourth column:} the zoomed in sectors.}
\label{fig:single-apperanceM1}
\end{figure}

Let us firstly analyze the rings and images of the  hairy Schwarzschild black hole with single photon sphere, and check the {effects} of the hairy parameters. The results depicted in  FIG.\ref{fig:single-apperanceM1} show significantly different behaviors for different hairy parameters. We discuss the main properties.  (i) The central dark region, i.e, the shadow, is smaller for larger deviation parameter $\alpha$,  but it becomes larger for stronger hairy charge $l_o$. This means that in certain cases, $l_o$ may balance the deviation of hairy black hole image from Schwarzschild black hole. These properties for shadow are {held} for other accretion model, because the critical impact parameter or the critical curve is mainly determined by the geometry itself, independent of the surroundings. (ii) The optical appearance image with $\alpha=2~\&~l_o=0.6$ is similar as that for Schwarzschild black hole \cite{Gralla:2019xty}, namely,  the direct intensity dominates the total luminosity under a bright ring (originated from direct intensity) of radiation, which encloses a thinner and dimmer ring contributed by the lensed ring intensity and ends in an even thinner ring originated from the photon ring intensity. When we decrease (increase) $l_o$ ($\alpha$), the three intensities could mix with each other such that the brightness origination of the total observed intensity at each $b$ could be completely different from the Schwarzschile (-like) cases. For $\alpha=2~\&~l_o=0.2$  due to the mixture of the direct intensity and lensed right intensity,  inside the bright ring  (originated from direct intensity and lensed right intensity) of radiation, there exists a wide bright region  enclosing a ring contributed from the direct intensity and then ending in a thinner ring originated from the photon ring intensity. For $\alpha=3~\&~l_o=0.2$, the direct intensity runs into the left side of photon ring intensity, so it contributes to the inner-most ring in the image. In a word, the radius and brightness contributions of the  light rings both closely depend on the hairy parameters. However, the optical appearance for hairy Schwarzschild black hole for certain hairy parameters could be the same as that for the Schwarzschild black hole, implying the potential degeneracy in the images, because $\alpha$ and $l_o$ may counteract each other's effects.

We move on to  diagnose the image of hairy Schwarzschild black hole with double photon spheres by the choice of  $\alpha=6.6~\&~l_o=0.9$.  The presence of a second photon sphere induces new ray sources in the region $b_{ph1} \leq b \leq b_{ph2} $, so it should introduce new ingredients in the optical appearances. The rich features can be seen in  FIG.\ref{fig:double-apperanceM1}, where  the inner edge of the disk in the emitted luminosity  locates at  $r_{isco}/M=1.0985$.
From the left plot, we see that in contract to the single photon sphere case, a new pole appears in each intensity, which leads to  the presence of additional peaks in the total observed intensity (in the middle plot). Thus, we see additional new light rings appearing in the inner region of the optical appearance image besides the usual three ones found in single photon sphere case.

%%%%%%%%%%%%%%
\begin{figure}[htbp]
\centering
{\includegraphics[width=5cm]{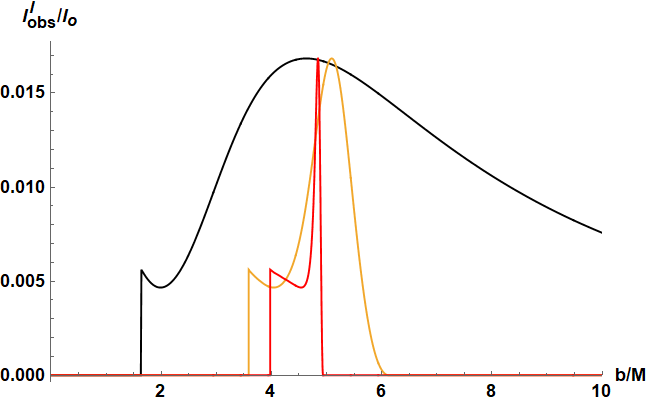}\hspace{2mm}
\includegraphics[width=5cm]{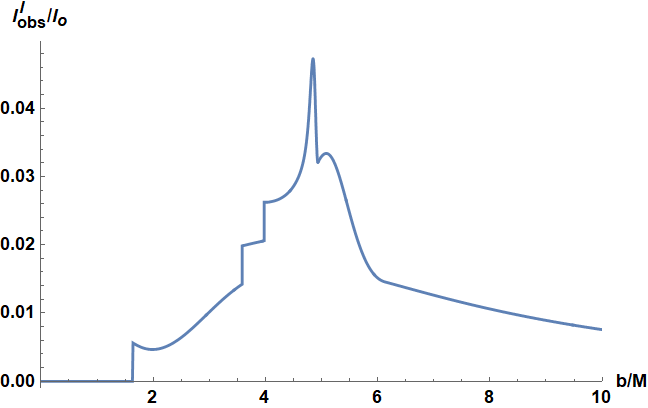}\hspace{2mm}
\includegraphics[width=4.5cm]{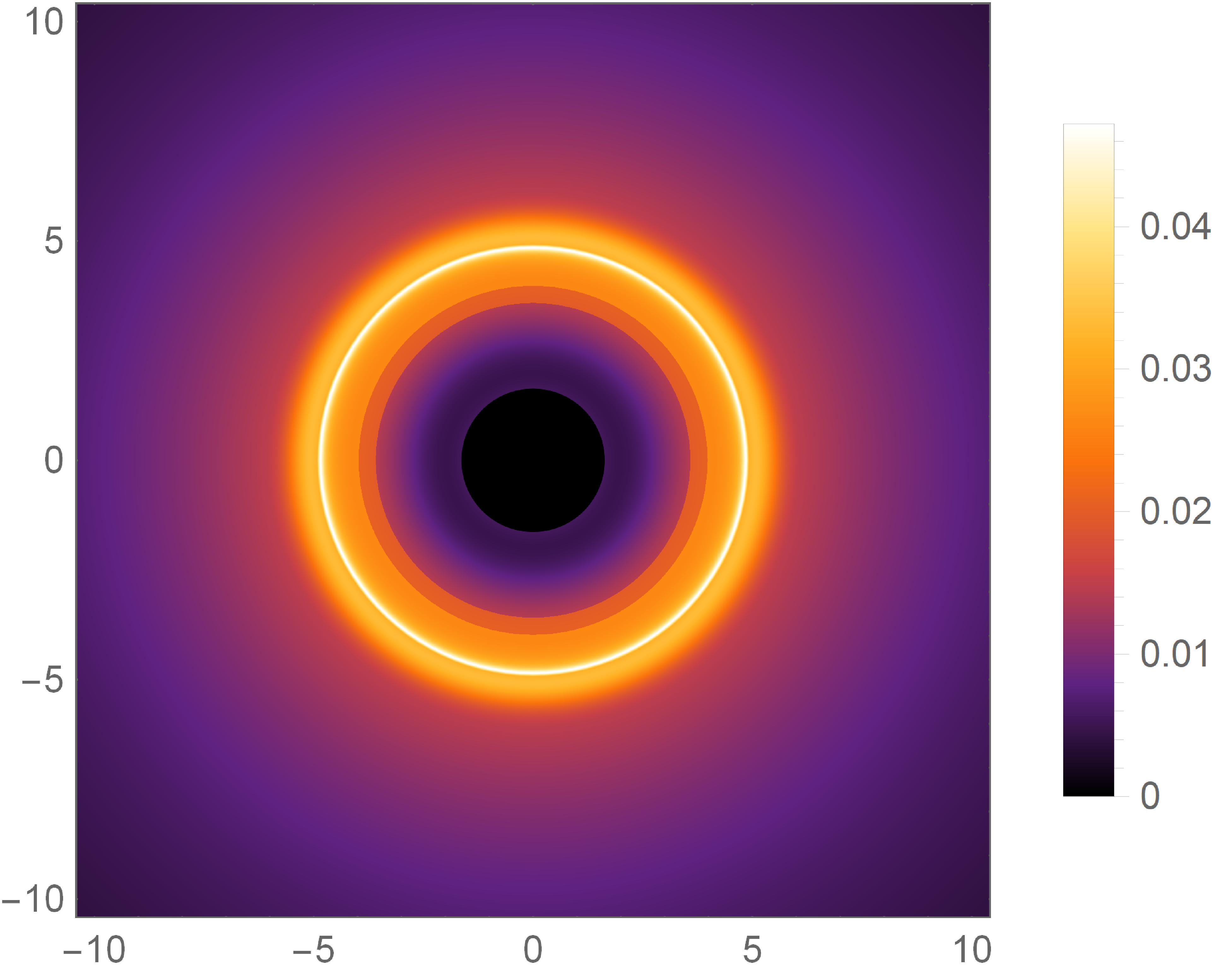}}
\caption{Observational appearances of hairy Schwarzschild black hole with double photon spheres ($\alpha=6.6$ and $l_o=0.9$) under the illumination of Model I \eqref{obser1}. \textbf{Left}: the different observed intensities originated from the first (black), second (gold) and third (red) transfer function in Eq. \eqref{eqtransfer} respectively. \textbf{Middle}: the total observed intensities $I_{obs}/I_{o}$ as a function of impact parameter $b$.  \textbf{Right}: the optical appearance: the distribution of observed intensities into two-dimensional disks.}
\label{fig:double-apperanceM1}
\end{figure}

\subsubsection{Model II}
In Model II, the emission is assumed to start from the photon sphere $r_{ph}$, and decay suppressed by the third power
\begin{figure}[H]
\begin{minipage}{0.5\linewidth}
\begin{eqnarray}
        I_{em}^{II}(r)=
        \begin{cases}
            I_o\frac{1}{[r-(r_{ph}-1)]^3}                 ~~~~~~~~~~~~& r>r_{ph}\\
            \\
            0,                                            ~~~~~~~~~~~~& r\leq r_{ph}
        \end{cases} ,  ~~~~~
    \label{obser2}
\end{eqnarray}
\end{minipage}
\hfill
\begin{minipage}{0.5\linewidth}
 \centerline{\includegraphics[width=1.5in]{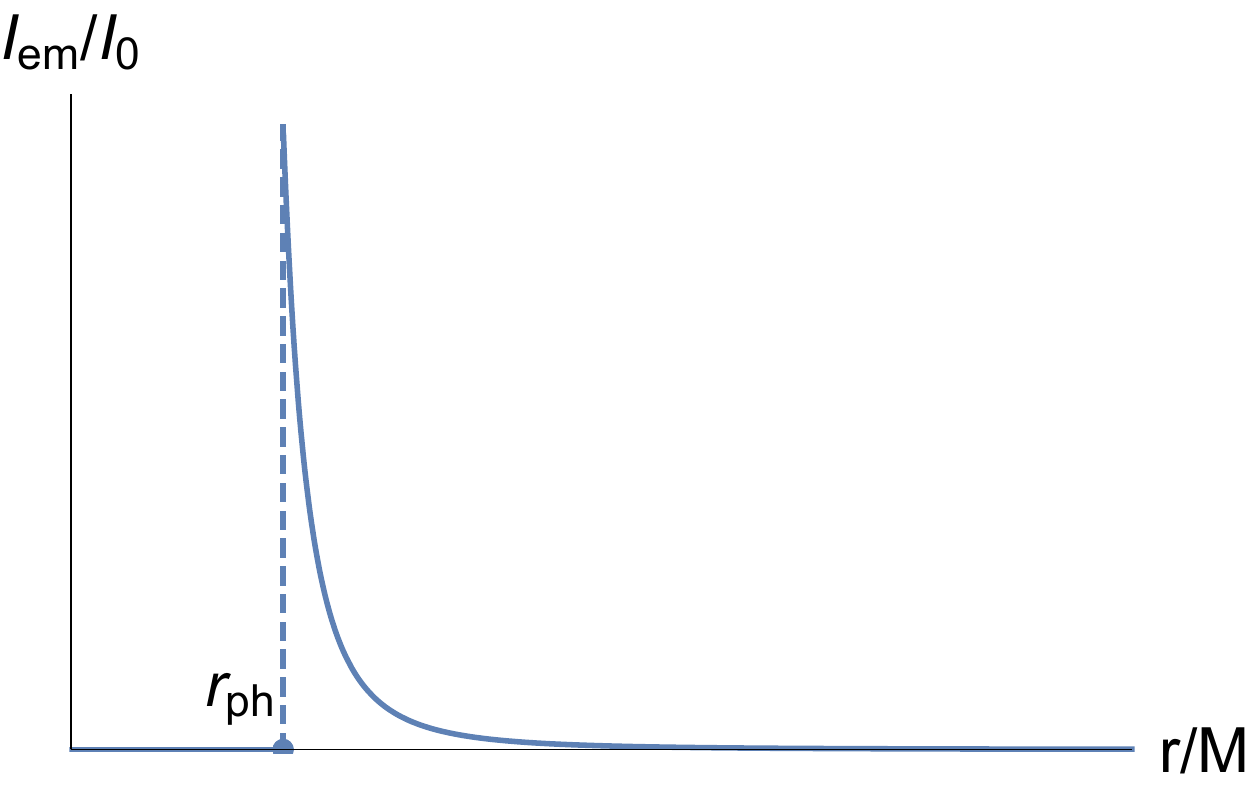}}
\end{minipage}
\end{figure}
The results are depicted in  FIG.\ref{fig:single-apperanceM2} (single photon sphere) and FIG.\ref{fig:double-apperanceM2} (double photon spheres). It is obvious from the left plots that this accretion construction  allows the direct intensity  to
cross  the critical impact parameter region due to the gravitational redshift, and becomes the dominant contribution there. As the
impact parameter increases, we will see the combinations among/beween the photon ring, lensed ring, and direct intensities into the total observed intensity, and the direct intensity with radiation dominates again as we further increase the impact parameter. Therefore, as shown in the right plot of  FIG.\ref{fig:single-apperanceM2},  the optical appearance of hairy Schwarzschild black hole with single photon sphere illuminated by Model II {is very similar to that for  Schwarzschild black hole \cite{Gralla:2019xty,Wang:2023vcv,Fathi:2023hwe,Guerrero:2022qkh}}. The general feature is that a dark shadow is surrounded by a wide region of luminosity enclosing two bright rings, and the intensity of the inner ring is contributed by the direct intensity while the outer ring is the joint result of all three intensities. In addition, the decreasing (increasing)  of $l_o$ ($\alpha$) will broaden the  wide region of luminosity but reduce the brightness of the light rings.

%%%%%
\begin{figure}[htbp]
\centering
\subfigure[\, $\alpha=2,l_o=0.2$]
{
  \includegraphics[width=5cm]{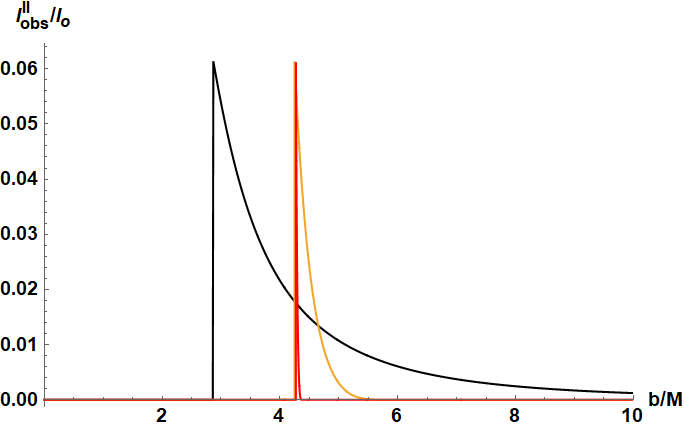}\hspace{2mm}
      \includegraphics[width=5cm]{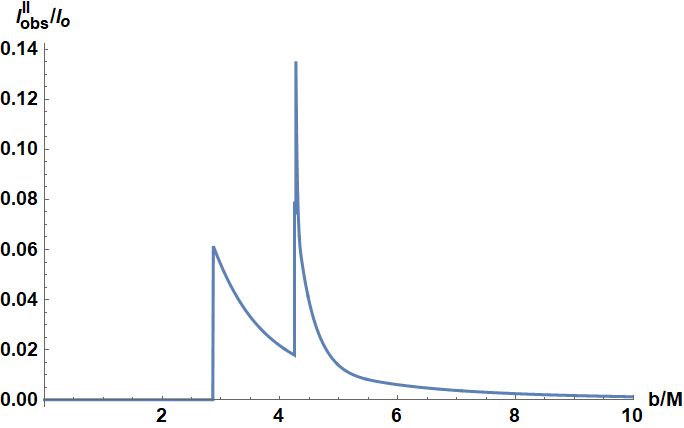}\hspace{2mm}
      \includegraphics[width=4.5cm]{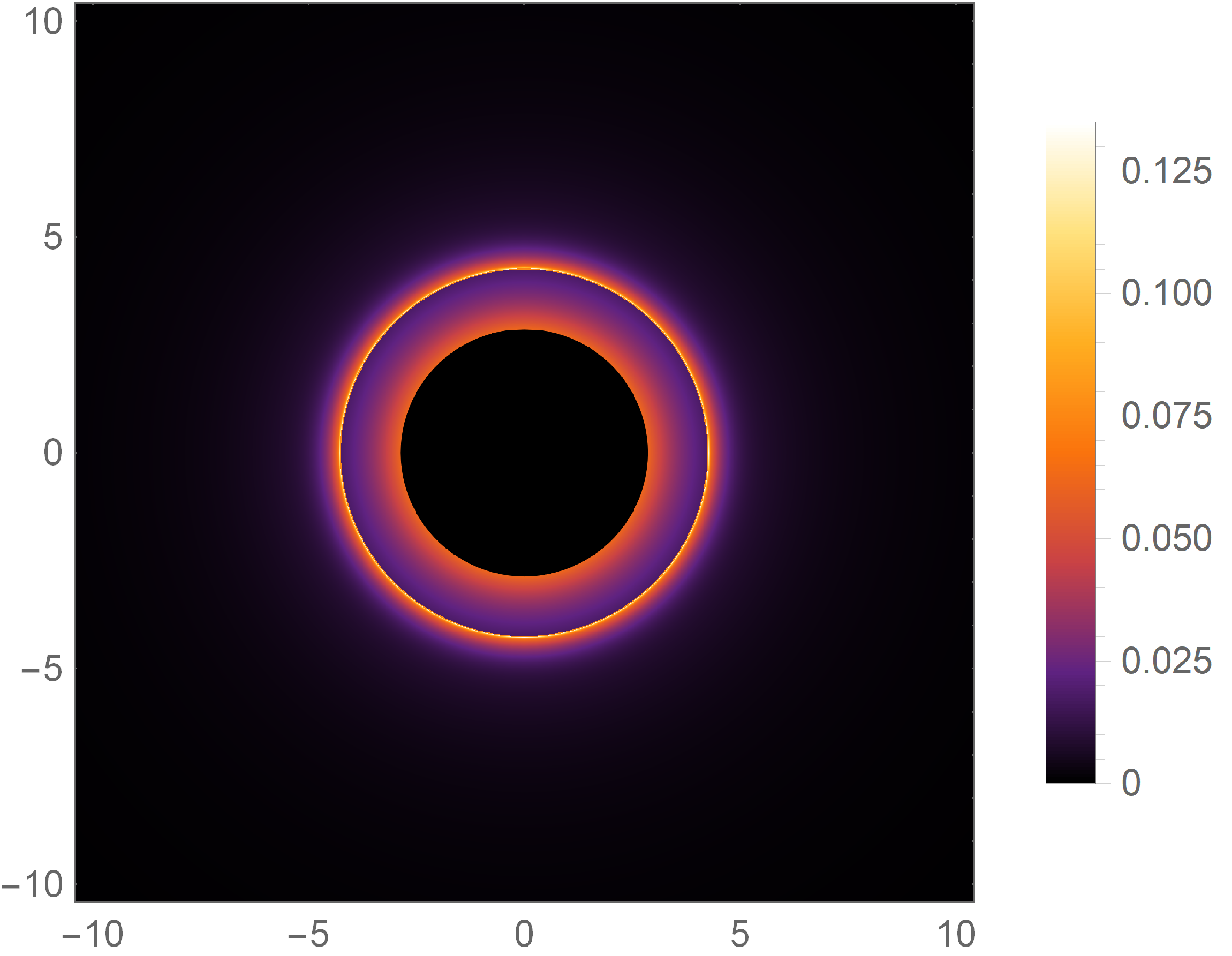}
\label{fig:single-apperanceM2-1}
}
\subfigure[\, $\alpha=2,l_o=0.6$]
{
      \includegraphics[width=5cm]{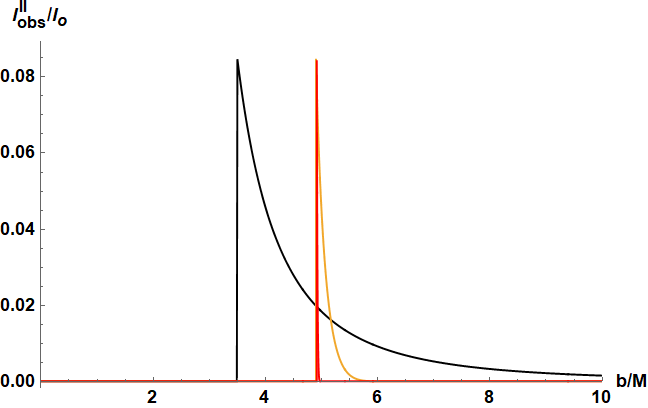}\hspace{2mm}
\includegraphics[width=5cm]{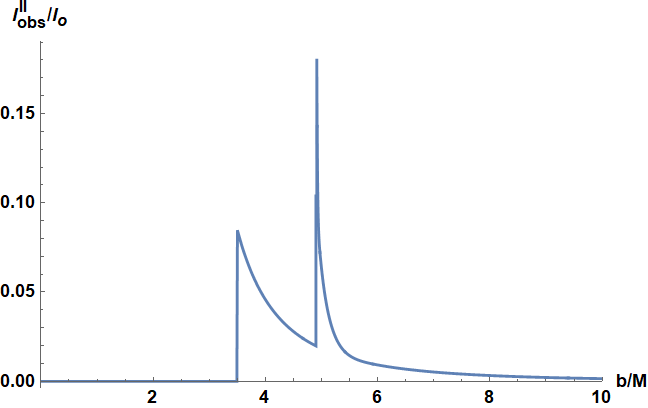}\hspace{2mm}
\includegraphics[width=4.5cm]{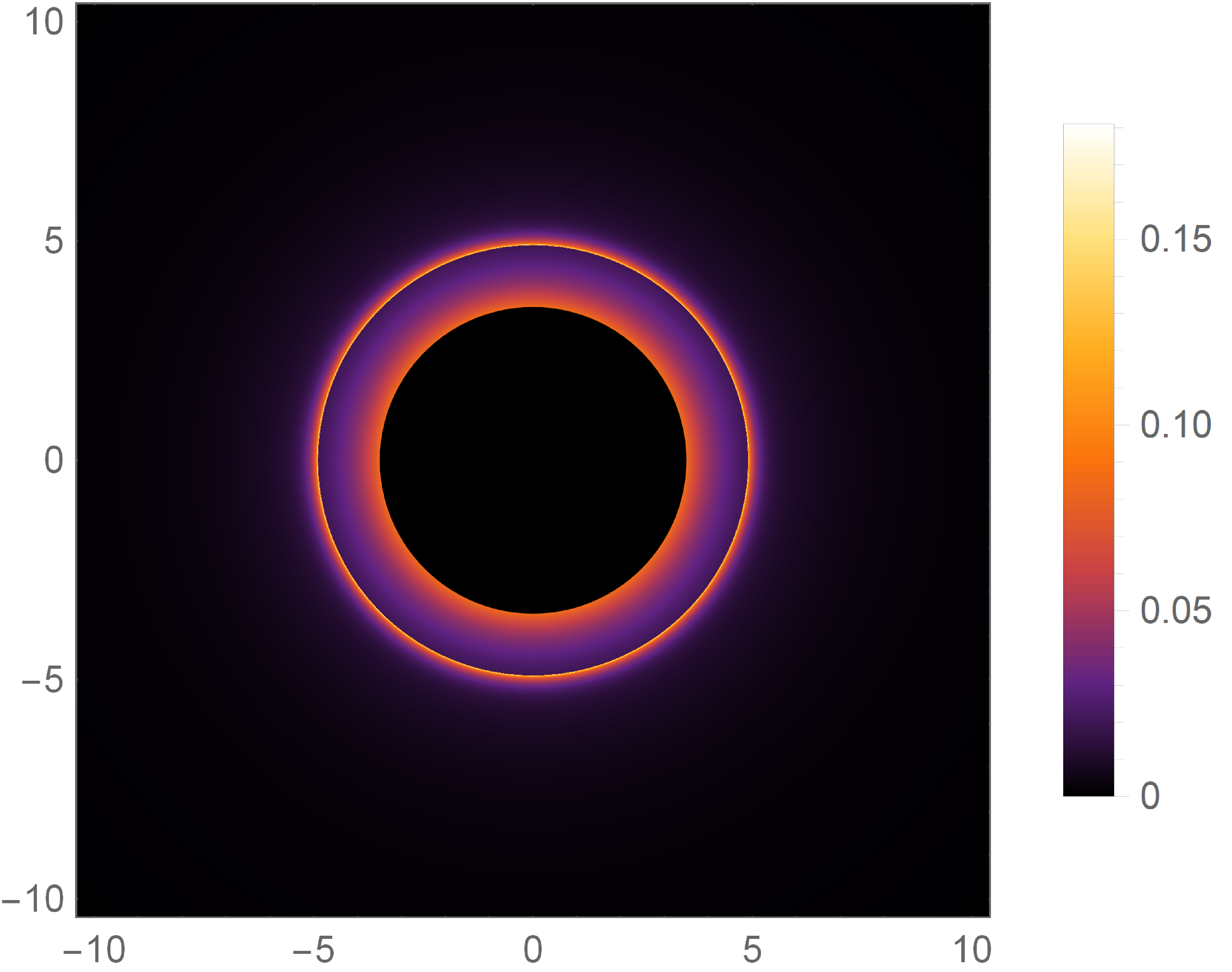}
\label{fig:single-apperanceM2-2}
}
\subfigure[\, $\alpha=3,l_o=0.2$]
{
\includegraphics[width=5cm]{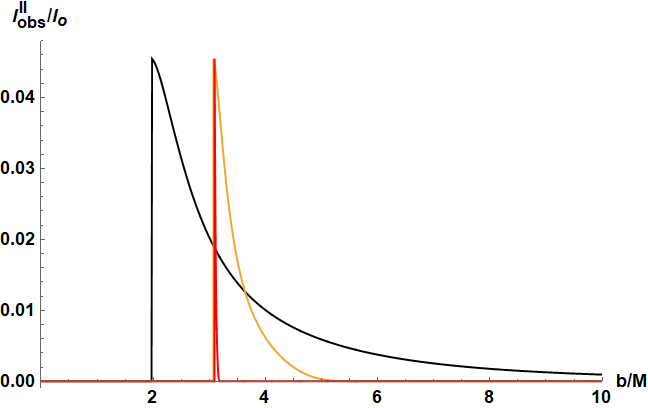}\hspace{2mm}
\includegraphics[width=5cm]{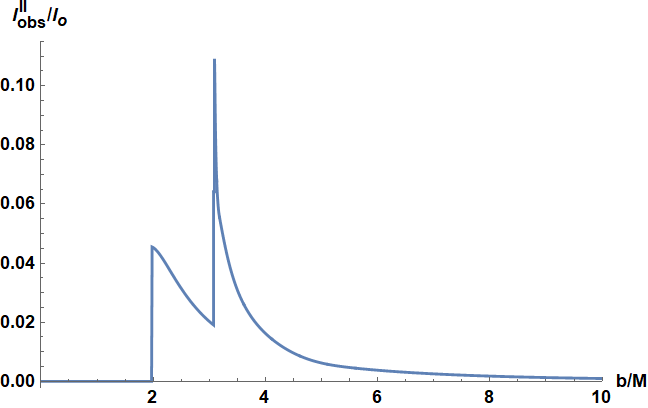}\hspace{2mm}
\includegraphics[width=4.5cm]{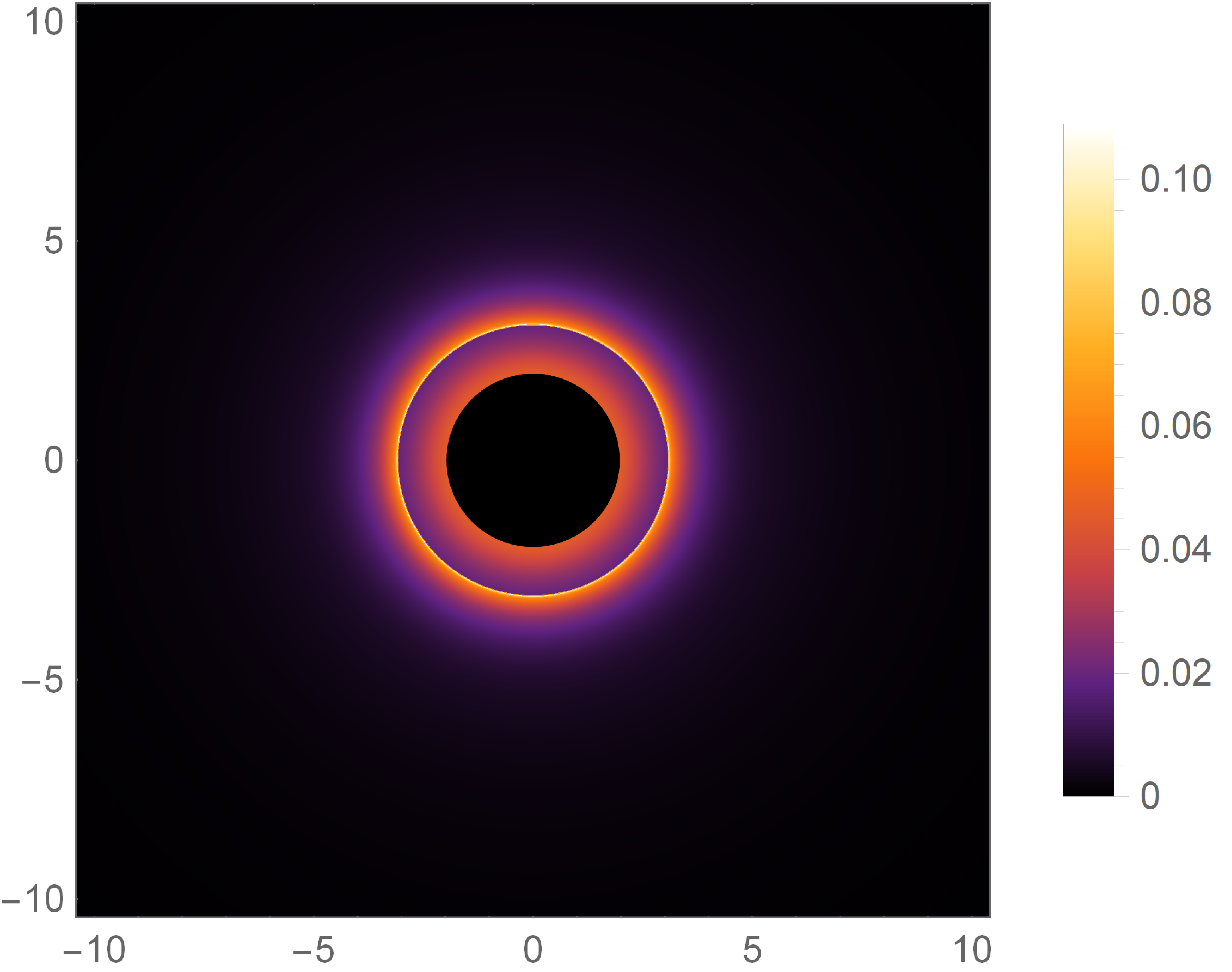}
\label{fig:single-apperanceM2-3}
}
\caption{Observational appearances of the Model II \eqref{obser2} of a thin disk for different $\alpha$ and $l_{o}$ whit $M=1$. \textbf{Left}: the different observed intensities originated from the first (black), second (gold) and third (red) transfer function in Eq.\eqref{eqtransfer} respectively. \textbf{Middle}: the total observed intensities $I_{obs}/I_{o}$ as a function of impact parameter $b$. \textbf{Right:} the optical appearance: the distribution of observed intensities into two-dimensional disks.}
\label{fig:single-apperanceM2}
\end{figure}

\begin{figure}[htbp]
\centering
{\includegraphics[width=5cm]{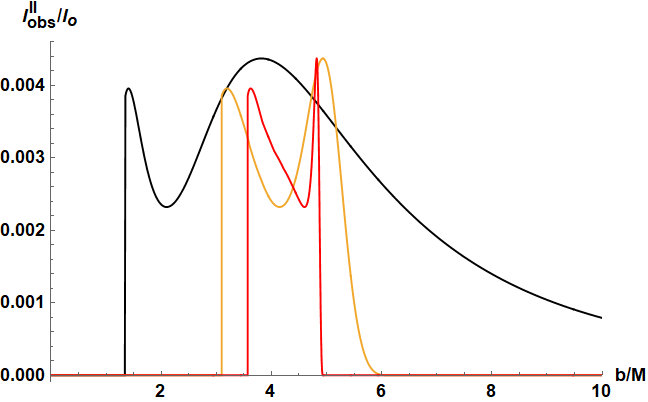}\hspace{2mm}
\includegraphics[width=5cm]{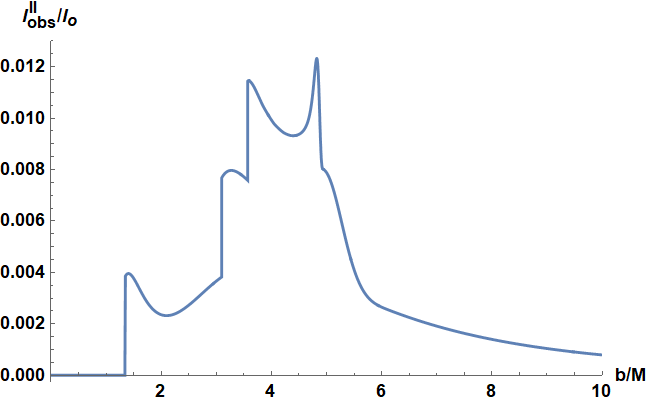}\hspace{2mm}
\includegraphics[width=4.5cm]{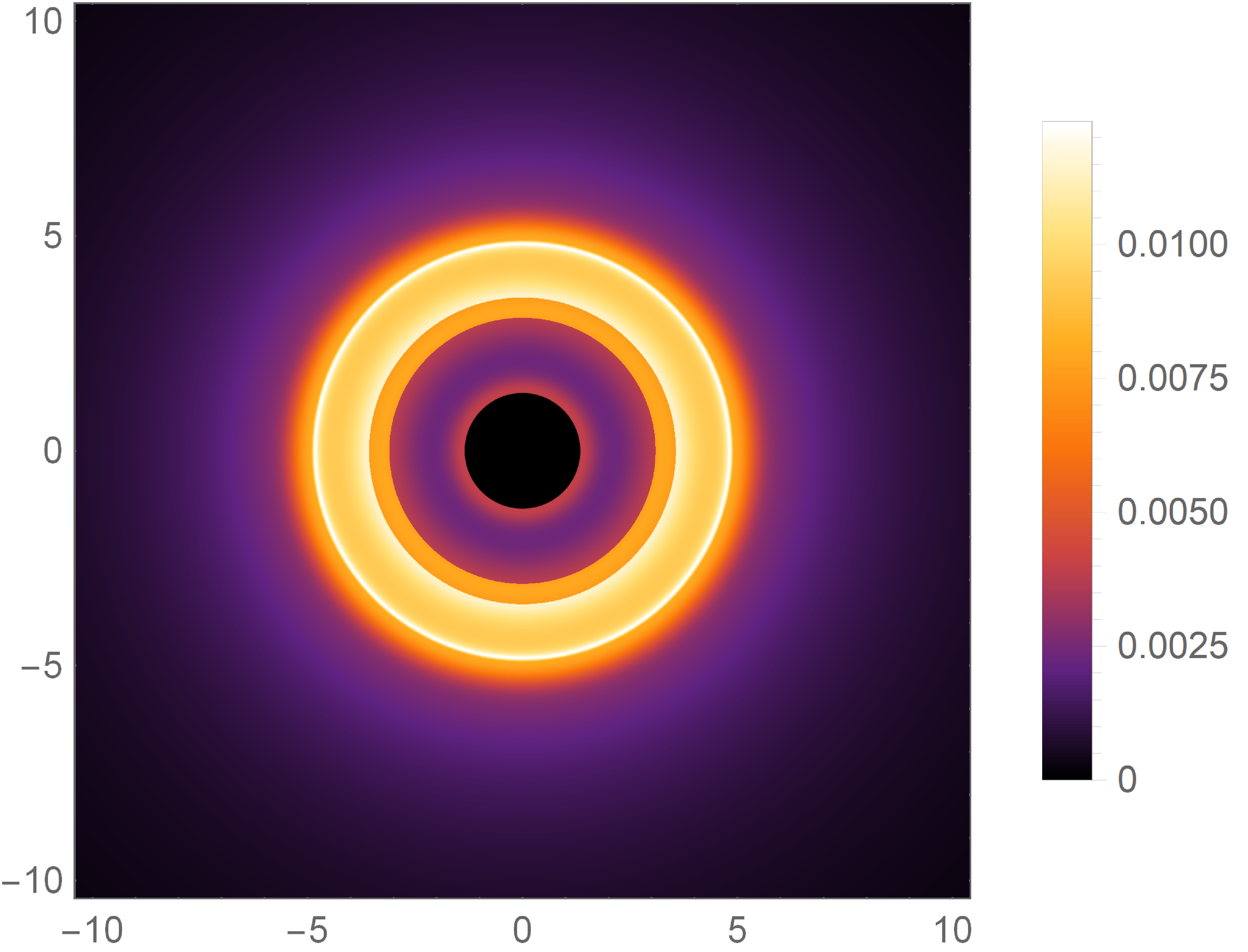}}\\
\caption{Observational appearances of hairy Schwarzschild black hole with double photon spheres ($\alpha=6.6$ and $l_o=0.9$) under the illumination of Model II \eqref{obser2}. \textbf{Left}: the different observed intensities originated from the first (black), second (gold) and third (red) transfer function in Eq.\eqref{eqtransfer} respectively. \textbf{Middle}: the total observed intensities $I_{obs}/I_{o}$ as a function of impact parameter $b$. \textbf{Right:} the optical appearance: the distribution of observed intensities into two-dimensional disks.}
\label{fig:double-apperanceM2}
\end{figure}

For the hairy Schwarzschild black hole with double photon spheres depicted in FIG. \ref{fig:double-apperanceM2},  the observed  intensities and their contributions' origination at small or large impact parameters are similar to that in single photon sphere case. However, in the intermediate impact parameter region, a second photon sphere again causes a new spike in each intensity and so additional peaks appear  in the total observed intensity. Therefore, the wide region of luminosity in the optical appearance could enclose more than two rings,  comparing to that in single photon region case.

\section{Shadows and rings with static thin spherical accretion} \label{sec:Static spherical accretions}

In this section, we will investigate the rings and images of hairy Schwarzschild  black hole surrounded by an optically and geometrically thin static  accretion
 with spherically symmetric. Thus, the specific intensity $I(\nu_o)$ observed by a distant observer at $r=\infty$ (measured in erg s$^{-1}$ cm$^{-2}$ str$^{-1}$ Hz$^{-1}$ ) radiated by the static accretion  can be obtained by integrating the specific emissivity along the photon path $\gamma$ \cite{Jaroszynski:1997bw,Bambi:2013nla}
\begin{equation}
I(\nu_o)=\int_{\gamma} g^3 j_e(\nu_e)dl_{prop},\label{eq-Intensity}
\end{equation}
where $g=\nu_o/\nu_e=f(r)^{1/2}$ is the redshift factor,  $\nu_o$ and $\nu_e$ are the observed photon frequency and the emitted photon frequency respectively. $j_e(\nu_e)$ is the emissivity per unit
volume in the rest frame and we will set {$j_e(\nu_e)\propto \delta(\nu_r-\nu_e)/r^2$ as usual \cite{Bambi:2013nla}, where $\nu_r$ is the emitter’s rest-frame frequency}. $dl_{prop}$ is the infinitesimal proper length,
\begin{equation}
dl_{prop}=\sqrt{\frac{1}{f(r)}dr^2+r^2d\phi^2}=\sqrt{\frac{1}{f(r)}+r^2\left(\frac{d\phi}{dr}\right)^2}dr,
\end{equation}
where the formula of $d\phi/dr$ can be read off from \eqref{trajectory-light-ray}.
Then, integrating
Eq.(\ref{eq-Intensity}) over all the observed frequencies, we get the total intensity observed by a distant observer
\begin{equation}
I_{obs}=\int_{\nu_o} I(\nu_o)d\nu_o=\int_{\nu_e}\int_\gamma g^4 j_e(\nu_e)dl_{prop} d\nu_e=\int_{\gamma}\frac{f(r)^{2}}{r^2}\sqrt{\frac{1}{f(r)}+r^2\left(\frac{d\phi}{dr}\right)^2}dr.\label{eq-Intensity2}
\end{equation}
It is obvious that the observed intensity depends on the radial distance $r$ and the impact parameter $b$.

\begin{figure}[htbp]
\centering
\subfigure[\, specific intensity $I_{obs}$]
{\includegraphics[width=4.5cm]{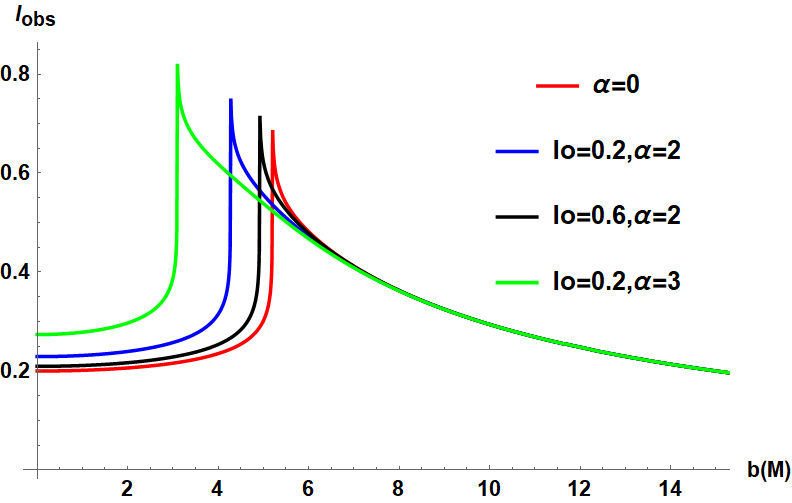}\label{fig:intensity-spherical-1}}\hspace{2mm}
\subfigure[\, $\alpha=2, l_o=0.2$]
{\includegraphics[width=3.2cm]{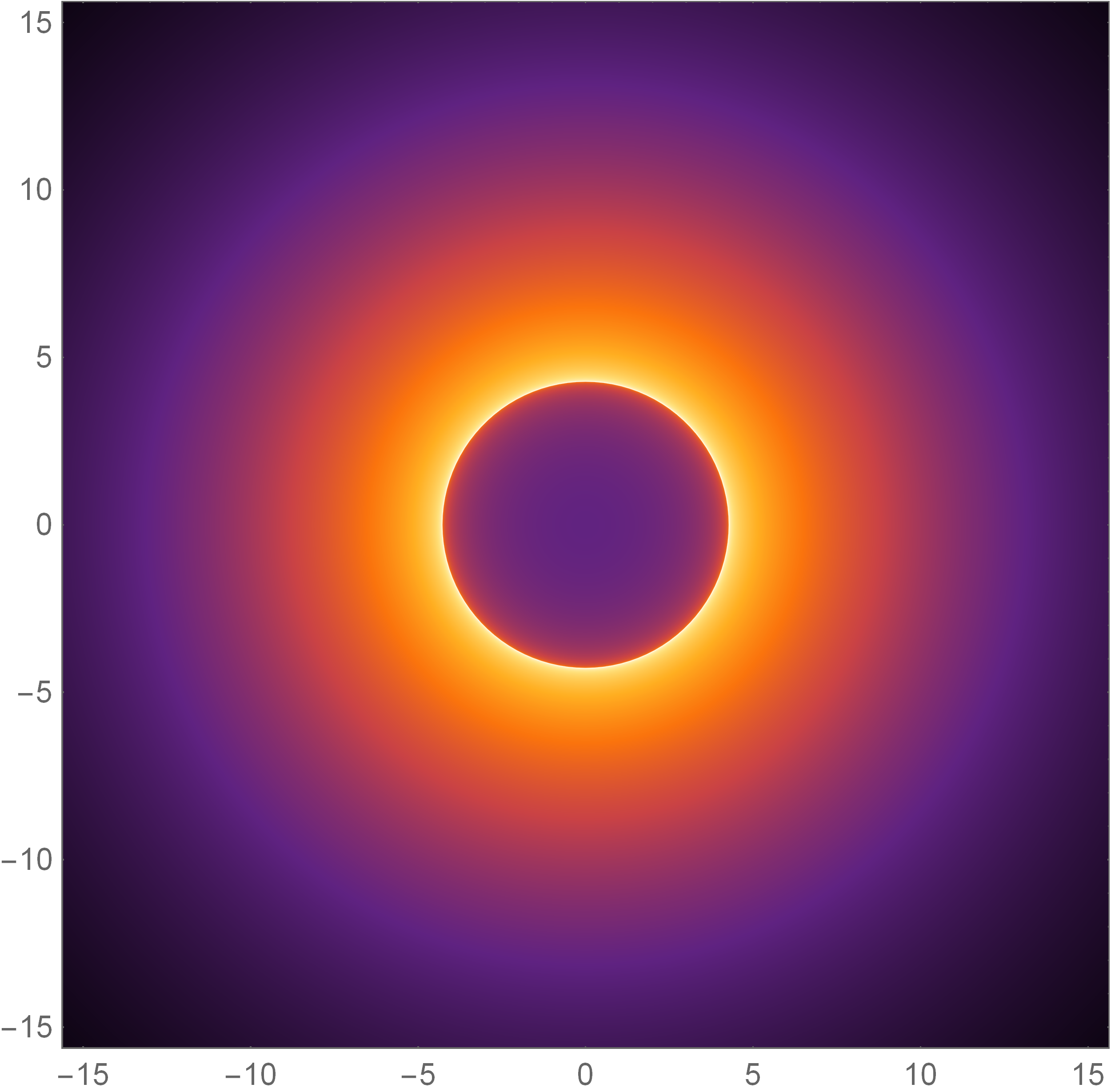}\label{fig:shadow-spherical-1-1}}\hspace{2mm}
\subfigure[\, $\alpha=2, l_o=0.6$]
{\includegraphics[width=3.2cm]{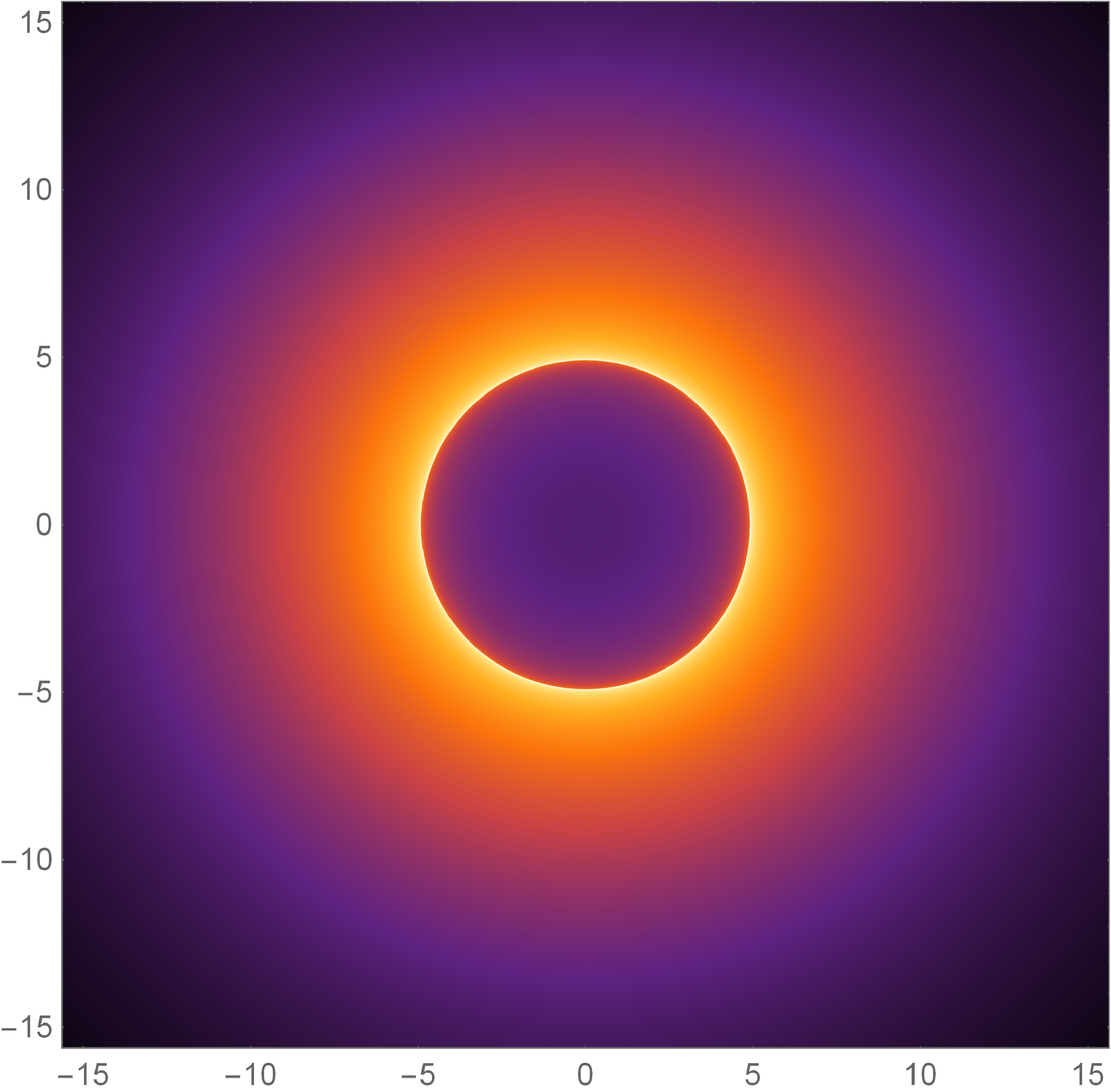}\label{fig:shadow-spherical-1-2}}\hspace{2mm}
\subfigure[\, $\alpha=3, l_o=0.2$]
{\includegraphics[width=4cm]{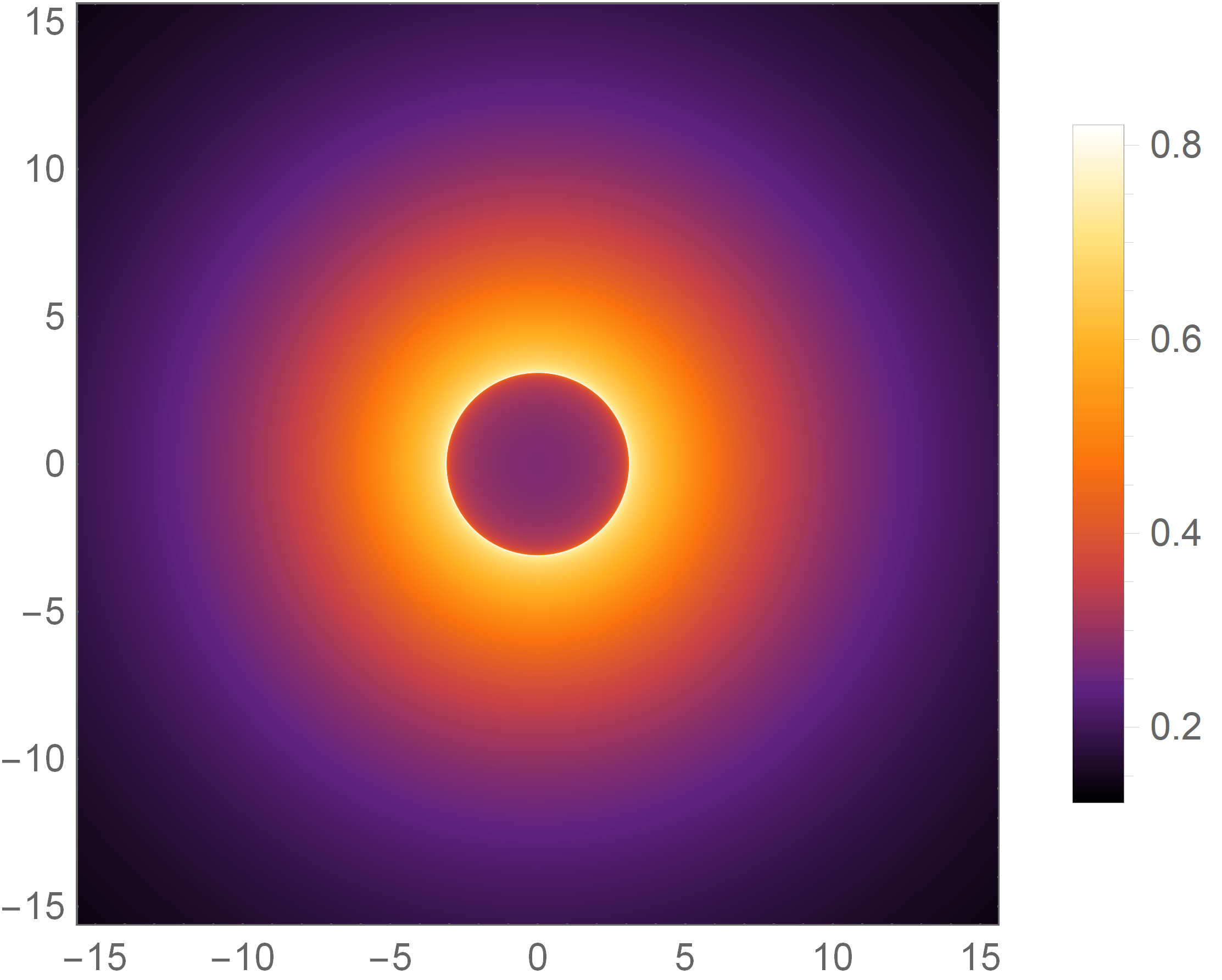}\label{fig:shadow-spherical-1-3}}
\caption{{\bf (a)}: The specific intensity $I_{obs}$ with respect to different parameters $\alpha$ and $l_o$.
{\bf (b)-(d)}: the black hole images with static spherical accretions for different $l_o$ and $\alpha$. Here we fix $M=1$.}
\label{fig:spherical-singleFS}
\end{figure}

\begin{figure}[htbp]
\centering
{\includegraphics[width=6.5cm]{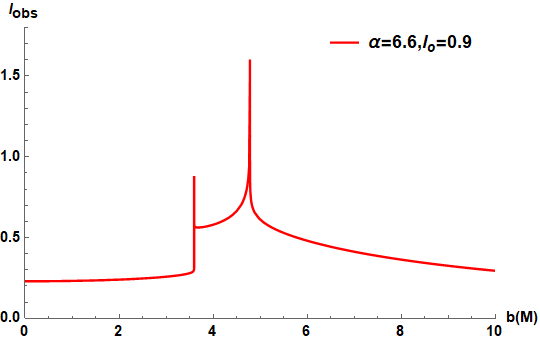}}\hspace{1cm}
{\includegraphics[width=5.5cm]{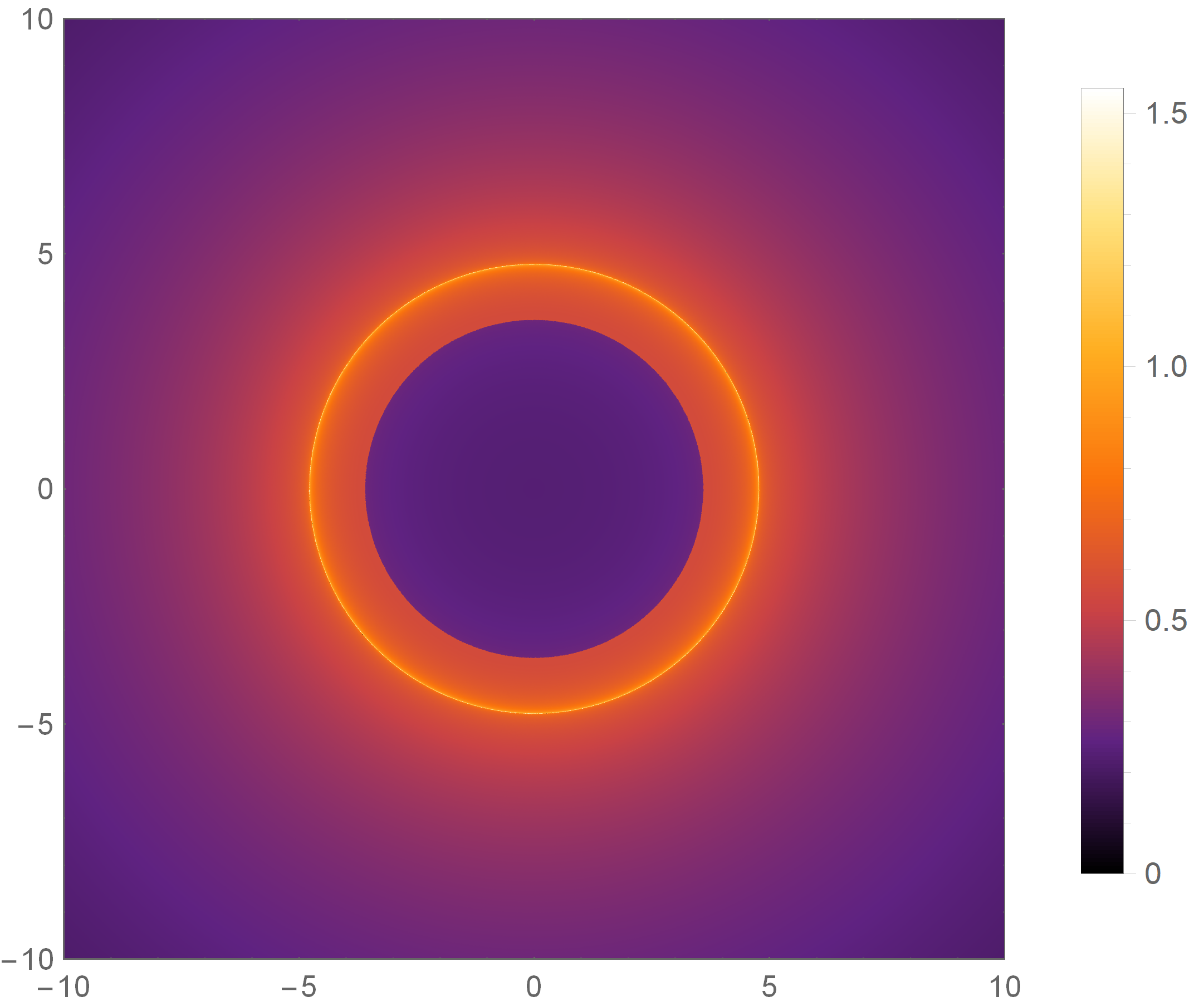}}\\
\caption{The specific intensity $I_{obs}$ of static spherical is given in the left panel. And the black hole image with double photon spheres static spherical accretion is given in the right panel. Here we fix $M=1$.}
\label{fig:spherical-doubleFS}
\end{figure}

The observed intensity for the hairy Schwarzschild black hole with single photon sphere is depicted in FIG.\ref{fig:intensity-spherical-1}, which shows that in each case, there is a peak indicating a bright ring in the image as found in Schwarzschild black hole.  As $l_o$ ($\alpha$) decreases (increases), the observed intensity including its peak will be enhanced, which means that the corresponding luminosity  of the black hole becomes brighter. Besides, for smaller (larger)  $l_o$ ($\alpha$),  the central region inside the bright ring shrinks while the bright luminosity region with  radiation region becomes wider. These observations again imply the competition of $l_o$  and $\alpha$ in their effects on the optical appearances of hairy Schwarzschild black hole. The picture described above is clear to see in the images of the black hole, shown in FIG. \ref{fig:shadow-spherical-1-1}-FIG.\ref{fig:shadow-spherical-1-3} which exhibit the distribution of total observed intensity in two-dimensional plane.

We move on to check the image of the hairy Schwarzschild black hole with double photon spheres, of which the observed intensity  is depicted in FIG. \ref{fig:spherical-doubleFS}. The key feature in this case is the appearing of two peaks in the total observed intensity (the left plot), indicating that two bright rings will be observed in the optical appearance image (the right plot). The additional
bright ring is again introduced by the second photon sphere, which allows the related black hole to be distinguished from the one with single photon sphere.

\section{Closing remarks}\label{sec:conclusion}
Studying the validity or violations of the no-hair theorem is one of the powerful ways to test gravity, and this subject becomes more fundamental when we refer to the existence of additional matter fields. The remarkable images of black holes published  by the EHT collaborators  open a new epoch of testing gravity in the strong field regime, including the probe of the validity or violations of  no-hair theorem. In this paper, we firstly analyzed the light rays around a hairy Schwarzschild black hole, which is a solution to the extended theory of general relativity including surrounding matter sources. Based on this, we then employed the ray tracing method to explore the shadows, rings and optical images of the hairy black hole when it is illuminated by static and (optically and geometrically) thin accretions. We found that comparing to the Schwarzschild black hole, the hairy parameters have significant influences in the distributions of light rays around the hairy black hole, such that the rings and images exhibit richer features.

Depending on the hairy parameters ($\alpha$ and $l_o$), the radial effective potential of the photon could have one or two maximum values. Though the configurations of the effective potential are diversified, we pointed out that for a far distance observer at north pole, the observational features of the hairy Schwarzschild black hole could be well reflected by two types of configurations, re-denoted by the hairy Schwarzschild black hole with single photon sphere and double photon spheres, respectively. In addition, the hairy charge $l_o$ enhances the event horizon, photon sphere, critical impact parameter and the borders of impact parameter for direct, lensed ring and photon ring emissions (see Table \ref{table01}). While, increasing the deviation parameter $\alpha$ suppresses all those values (see Table \ref{table02}). These competitive effects could interpret the potential degeneracies in the images between the hairy Schwarzschild black hole and Schwarzschild black hole, when they are illuminated by accretions.

After the impact parameter region was determined, we considered two toy models of light sources as static thin accretions disk, of which the inner edge locates at the ISCO for timelike object and the photon sphere, respectively. We firstly analyzed the first three transfer functions which connect $m-$hitting points on the disk with the impact parameter of the light ray. We found that for hairy Schwarzschild black hole with single photon sphere, the brightness contributions from the second and third transfer functions are puny comparing to the first transfer function because they  demagnify sharply (FIG.\ref{fig:transfer function}\blue{a}).
While for the hairy Schwarzschild black hole with double photon spheres,  the widths of the second and third transfer functions have been significantly widened, and in certain region, their demagnification even smaller than that for first transfer function (FIG.\ref{fig:transfer function}\blue{b}). By collecting the transfer functions, we calculated the corresponding observed intensities respectively originated from the direct, lensed ring and photon ring intensity. And then we evaluated their resultant total observed intensity for both accretion models,  from which we transfer to the optical appearance image in the observer's plane.  Our observations can be summarized as follows.
\begin{itemize}
  \item The images of the hairy Schwarzschild black hole are very different depending on the emission model of the accretion disk, which also happens for Schwarzschild black hole. However, the central dark region, i.e, the shadow, is smaller (larger) for larger $l_o$ ($\alpha$). These properties for shadow are independent of the disk models, because the critical impact parameter or the critical curve is mainly determined by the geometry itself rather than the surroundings.

  \item For the hairy Schwarzschild black hole with single photon sphere, we can always observe three light rings in the images for model I (FIG.\ref{fig:single-apperanceM1}), however, the hairy parameters seriously affect the position and width of the light rings. So we may observe completely different optical appearances from that of Schwarzschild black hole, but due to the competitive effects of the two parameters, we can also obtain the degeneracies of images between hairy Schwarzschild black hole and Schwarzschild black hole.

  While under the illumination of Modle II, we found that the optical appearance of hairy Schwarzschild black hole with single photon sphere is very similar to that for Schwarzschild black hole. Namely, a dark shadow is surrounded by a wide region of luminosity enclosing two bright rings (FIG.\ref{fig:single-apperanceM2}). However, it was observed that decreasing (increasing) $l_o$ ($\alpha$) could broaden the  wide region of luminosity but obfuscate the brightness of the light rings.

  \item For the hairy Schwarzschild black hole with double photon spheres, it is obvious that a second photon sphere could introduce additional peaks in the total observed intensity in both accretion disk models (FIG.\ref{fig:double-apperanceM1} and FIG.\ref{fig:double-apperanceM2}). Thus, when compared to the case of a single photon sphere, additional new light rings were observed in the images of those black holes, making them distinguishable.
\end{itemize}

Finally, we investigated the rings and images of the hairy Schwarzschild black hole under the illumination of a static, spherically symmetric thin accretion. We also found that $l_o$  and $\alpha$ have competitive effects on shadow size, brightness of ring in the optical appearance image (FIG.\ref{fig:spherical-singleFS}). This again could lead to the potential degeneracies between the black holes with and without hair. Moreover, under the spherical accretion, we observed two bright rings in the optical appearance of hairy Schwarzschild black hole with double photon spheres (FIG.\ref{fig:spherical-doubleFS}). This also provides an alternative tool to testify black holes with double photon spheres.

\begin{acknowledgments}
This work is partly supported by  Natural Science Foundation of Jiangsu Province under Grant No.BK20211601, Fok Ying Tung Education Foundation under Grant No.171006, the Postgraduate Research \& Practice Innovation Program of Jiangsu Province under Grant No. KYCX22\_3452 and KYCX21\_3192, and Top Talent Support Program from Yangzhou University.
\end{acknowledgments}

\bibliography{ref}
\bibliographystyle{apsrev}

\end{document}